\documentclass[letterpaper,twocolumn,10pt]{article}
\usepackage{usenix2019_v3}

\usepackage{tikz}
\usepackage{amsmath}
\usepackage[english]{babel}
\usepackage{subfigure,hyphenat,enumitem}
\usepackage{algorithm}
\usepackage[noEnd]{algpseudocodex}
\usepackage{amsmath,hyperref}
\usepackage{tabularx,booktabs,tablefootnote}
\usepackage[labelformat=simple,skip=0pt]{subcaption}
\usepackage[normalem]{ulem}
\usepackage{listings}
\usepackage{xcolor}
\usepackage[utf8]{inputenc}
\usepackage{bm}
\usepackage{amssymb,xfrac}
\usepackage[most]{tcolorbox}

\definecolor{codegreen}{rgb}{0,0.6,0}
\definecolor{codegray}{rgb}{0.5,0.5,0.5}
\definecolor{codepurple}{rgb}{0.58,0,0.82}
\definecolor{backcolour}{rgb}{0.95,0.95,0.92}

\lstdefinestyle{mystyle}{
    backgroundcolor=\color{backcolour},   
    commentstyle=\color{codegreen},
    keywordstyle=\color{magenta},
    numberstyle=\tiny\color{codegray},
    stringstyle=\color{codepurple},
    basicstyle=\ttfamily\tiny,
    breakatwhitespace=false,         
    breaklines=true,                 
    captionpos=b,                    
    keepspaces=true,                 
    numbers=left,                    
    numbersep=5pt,                  
    showspaces=false,                
    showstringspaces=false,
    showtabs=false,                  
    tabsize=2
}

\usepackage{tabularx,wrapfig}
\usepackage[capitalize]{cleveref}
\usepackage{filecontents}

\DeclareCaptionLabelSeparator{emdash}{--- }

\newcommand{\parahead}[1]{\vspace*{1ex plus 0ex minus 0.25ex}\noindent{}{\bfseries #1}}
\newcommand{\subparahead}[1]{\vspace*{1ex plus 0ex minus 0.25ex}\noindent{}{\itshape #1}}

\setitemize{itemsep=0pt,topsep=3pt,parsep=0pt,partopsep=0pt,leftmargin=1.0em}
\setenumerate{itemsep=2pt,topsep=3pt,parsep=2pt,partopsep=0pt,leftmargin=1.5em}
\setlist{itemsep=2pt,parsep=2pt}

\newcommand{\parabreak}{\vspace*{0.5ex plus 0.5ex}\noindent}

\newcommand{\sysname}{\textsf{CellTwin}}
\newcommand{\sysnames}{\textsf{CellTwin's}}
\newcommand{\trafficmodel}{\textsf{Traffic Reconstructor}}
\newcommand{\schedulermodel}{\textsf{Neural Scheduler}}
\newcommand*\circled[1]{\tikz[baseline=(char.base)]{
            \node[shape=circle,draw,inner sep=1pt] (char) {#1};}}
\begin{document}
\date{}
\title{\Large \bf CellTwin: Reproducible Cellular Network Emulation via\\High Resolution, \textit{in situ} Measurement-Driven ML}
\author{
{\rm Haoran Wan}\\
Princeton University
\and
{\rm Yaxiong Xie}\\
University at Buffalo, SUNY
\and
{\rm Kyle Jamieson}\\
Princeton University
} 

\maketitle
\begin{abstract}
Current and future applications demand ultra-low latency and consistent throughput, yet frequently traverse 5G cellular networks, so must cope with volatile packet dynamics, as 5G base station schedulers dynamically react to user workloads and wireless channel conditions. The task of evaluating network algorithms in these environments is hamstrung by current tools: record-and-replay emulators sever the feedback interaction that exists between application end points and a commercial operator's proprietary 5G scheduler, while full-stack simulators rely on overly simplistic scheduling logic. To bridge this reality gap, we present \sysname{}, a high-fidelity, machine learning-based emulation framework that learns complex 5G scheduler resource allocation behaviors directly from extremely high-resolution network telemetry tools. The first emulator to handle multiple clients, \sysname{} utilizes machine learning to dynamically predict resource block allocations and modulation schemes based on instantaneous user buffer occupancy and channel states.  To capture realistic cross-user contention, a traffic reconstruction model inverts cellular network scheduling results to recover the underlying traffic patterns of uncontrolled background users.  Implemented as a high-performance Linux middlebox emulator, \sysname{} reduces emulation error relative to the state of the art for various network applications including but not limited to 55\% for web-page load time, 57\% for WebRTC encoder bit rate, and 51\% for cloud gaming packet one-way delay, providing an accurate, standardized testing ground for tomorrow's real-time interactive network protocols and applications.
\end{abstract}
\section{Introduction}

Driven by demand for ubiquitous, reliable connectivity and low latency, 
5G mobile networks are rapidly evolving to 
support a new generation of 
demanding, highly interactive applications: 
immersive cloud gaming \cite{wan_evolving_2024,ye_dissecting_2024,liDissectingStreamliningInteractive2025}, 
real-time video edge analytics 
\cite{10472093, 10.1145/3769102.3770618, 10.1145/3307334.3328589},
high-fidelity video conferencing \cite{yi_athena_2024,yiAutomatedCrossLayerRoot2025}, 
and embodied intelligence \cite{liLargeModelEnabled2025,SynergeticEmpowermentWireless} 
increasingly rely on 
5G to function in highly mobile environments. 

\begin{figure}
    \centering
    \includegraphics[width=\linewidth]{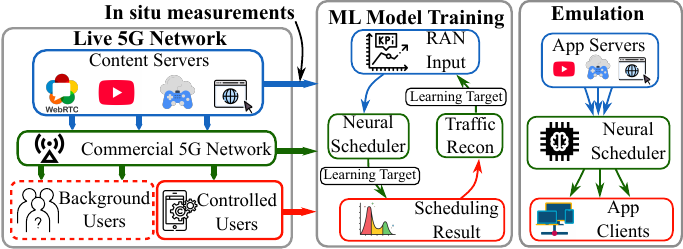}
    \caption{\sysname{} learns the scheduling behavior via \textit{in situ} measurement in the live 5G network for faithful emulation.}
    \label{fig:opener}
    \vskip -0.15in
\end{figure}

To support these workloads, such applications require sustained throughput 
and wired-level reliability 
\cite{chang:can-you-see-me, varvello:vca-performance-in-wild, yu:mustang}.
And while the widespread deployment of the
5G New Radio \textit{radio access network} (RAN) 
promises the massive bandwidth 
and ultra-low latency theoretically necessary to support these 
applications, cellular performance remains highly volatile in 
everyday deployment.
Plagued by channel fading, mobility-induced handovers, 
environmental obstructions, and complex base-station scheduling \cite{sentosaCellReplayAccurateRecordandreplay2025}, real-world
cellular channels still struggle to match the stable, consistent baseline
provided by Wi-Fi or wired networks
\cite{yiAutomatedCrossLayerRoot2025}.
Since the underlying physical layer cannot entirely mask this 
volatility, the burden of ensuring high performance and/or
quality of experience eventually falls on end-to-end network algorithms.
Foundational network protocols---congestion control \cite{246302} and 
\textit{adaptive bitrate} (ABR) algorithms \cite{211251}---must evolve 
to navigate the highly dynamic idiosyncrasies of modern 5G channels.

However, the perennial \cite{10.1145/972374.972386, 10.1145/3097766.3097772} and hard 
problem of reproducible wireless evaluation continues to bottleneck the
evolution of these network protocols. 
The gold standard, testing directly 
on live commercial cellular networks \cite{sentosaCellReplayAccurateRecordandreplay2025},
is notoriously time-consuming, strongly influenced by location, and 
fundamentally irreproducible. 
Conversely, researchers relying on existing network simulators \cite{nsnamNs3,system_srsran_2023} 
or trace-driven record-and-replay emulators \cite{netravali_mahimahi_2015,sentosaCellReplayAccurateRecordandreplay2025} 
often encounter a massive reality gap \cite{yuMagpieImprovingEfficiency2024,zhangArsenalUnderstandingLearningBased2021}
as these tools struggle to accurately replicate the complex, workload-dependent behaviors of modern RANs. 
This lack of accurate tooling frequently leads to biased evaluations: an ABR or congestion control algorithm that performs perfectly in a constrained simulator often degrades severely in the wild \cite{yuMagpieImprovingEfficiency2024}. 
If researchers cannot accurately and easily test how an algorithm handles a true 5G environment, algorithmic innovation stagnates.

We step back from these approaches, instead asking the following question: \textit{what basic design of 
a cellular network emulator would minimize the gap between emulation and reality?}
Inside the RAN, a scheduler maps the current \textit{UE state} (per-user buffer occupancy and channel status) to the dynamic bandwidth resource allocation at a \textit{slot} (half-millisecond in 5G) granularity, determining the instantaneous throughput for each user (UE).
This mapping is the root cause of throughput volatility and inter-UE contention patterns, so recreating this mapping faithfully is critical to narrow the gap.

\parabreak{}To achieve this goal, we present \sysname{}, an intuitive, high-fidelity, \textit{machine learning} (ML)-based evaluation framework built specifically to bridge the gap between theoretical algorithm design and real-world cellular performance. 
Unlike oversimplified simulators \cite{nsnamNs3,system_srsran_2023} or prior record-and-replay tools Mahimahi \cite{netravali_mahimahi_2015} and CellReplay \cite{sentosaCellReplayAccurateRecordandreplay2025}, \sysname{} learns the network's underlying resource allocation behavior directly from high resolution, \textit{in situ} wireless network measurements. 
\sysname{} combines passive \cite{wan_nr-scope_2024, xie_ng-scope_2022} and on-device telemetry tools \cite{liuSeeingFogEmpowering2025} for RAN metrics collection.
\sysname{} deploys randomized application flows with time-synchronized endpoints to induce diverse RAN buffer patterns and ensure reliable RAN buffer estimates.
To power this emulation environment, \sysname{} relies on two synergistic ML models. First, the \schedulermodel{} serves as the core emulation engine; it ingests the instantaneous buffer occupancies and \textit{channel quality indicators} (CQI) of all active users to dynamically generate resource schedules, faithfully mimicking a commercial 5G base station.
Second, to accurately capture cross-user contention within the emulator, the \trafficmodel{} inverts observable scheduling telemetry to deduce the hidden CQI and ingress traffic patterns of uncontrolled background users, reconstructing the cell's overall buffer dynamics. 
Together, these components forge a highly accurate emulator of the RAN, providing a standardized testing ground that enables rapid iteration of next-generation network algorithms.

\sysname{} is the first cellular network emulator tool that eschews
the extremes of record-and-replay based emulation on one hand and the use
of network simulators or open-source 5G stacks on the other. It is the 
first cellular emulator to accurately handle multiple users sharing the
same base station. Our implementation contributions include an automatic randomized data collection framework and a comprehensive suite of models representing a commercial base station. 
Specifically, \sysname{}'s high-resolution in situ telemetry dataset, and the models, alongside an experimental framework enabling plug-and-play algorithm evaluation are available at https://github.com/PrincetonUniversity/CellTwin.

Top-line results show that \sysname{} significantly outperforms state-of-the-art emulators in replicating true commercial 5G performance, including but not limited to reducing the ABR bit rate emulation errors from 7\% to below 1\%, and over 80\% reduction of congestion control RTT emulation errors, and a 55\% improvement in web-page load time emulation.
%

\parabreak{}The authors attest that this work raises no ethical issues.



\section{Primer: 5G New Radio}\label{s:background:5g}

\begin{figure}
    \centering
    \includegraphics[width=\linewidth]{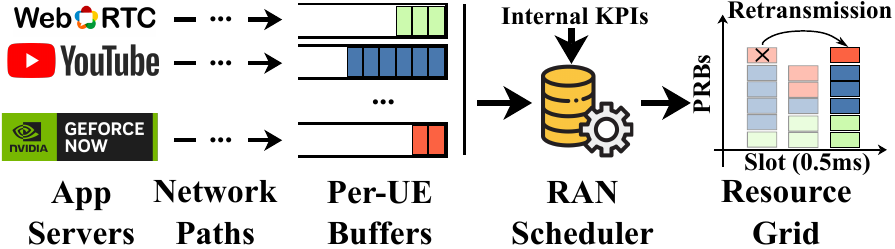}
    \caption{Cellular network schedules resources for users based on their buffer occupancy and internal KPIs (CQIs, QoS, etc.).}
    \label{fig:5g_background}
    \vskip -0.15in
\end{figure}

Viewed through a networking lens, a 5G base station is a highly dynamic bottleneck 
that manages a shared pool of wireless resources among multiple concurrent users, as shown in \cref{fig:5g_background}.
To isolate traffic and prevent direct interference between flows, the base station assigns 
and maintains one or multiple independent \textit{packet buffers} for each connected UE. 
The flow of data out of these individual buffers is governed by a central 
scheduling mechanism at the base station in the \textit{medium access control} (MAC) layer. 
Unlike wired links that service queues at a relatively constant rate, the 
5G scheduler dynamically allocates \textit{physical resource blocks}
(PRBs), the unit of communication granularity in time and frequency, to users, and 
assigns each PRB a \textit{modulation and coding scheme} (MCS) bit-rate
by continuously evaluating a complex, multidimensional set of inputs.\footnote{The 
total number of available PRBs is determined by the cell's configured frequency
bandwidth---\textit{e.g.}, the maximum number of 
PRBs in sub-6~GHz network is 273 (with 100~MHz bandwidth). 
MCS, indexed from 0-31, then 
determines the number of bits of data that can be carried by each PRB.}
Then the total data size delivered by the scheduling result at the given slot is called 
\textit{transport block size} (TBS).
Specifically, scheduling decisions are driven by the user's real-time 
physical channel status, their instantaneous buffer occupancy, and 
proprietary, vendor-specific \textit{key performance indicators} 
(KPIs) designed to balance overall cell throughput, fairness, 
and strict QoS targets. 
Consequently, the bandwidth experienced by an application 
is not a static property of the link, but rather the continuous output of 
this opaque, state-dependent scheduling logic.
Further compounding this variability is the base station's two-tiered
retransmission architecture, which utilizes rapid, low-level retries 
(Hybrid-ARQ) alongside \textit{radio link control} (RLC) link-layer 
recoveries to mask the inherent unreliability of the wireless medium. 
These mechanisms introduce highly variable packet delays that crucially depend 
on transient channel fluctuations and the incoming traffic 
workload.

 \begin{figure*}
    \begin{minipage}[b]{0.325\linewidth}
        \centering
        \includegraphics[width=\linewidth]{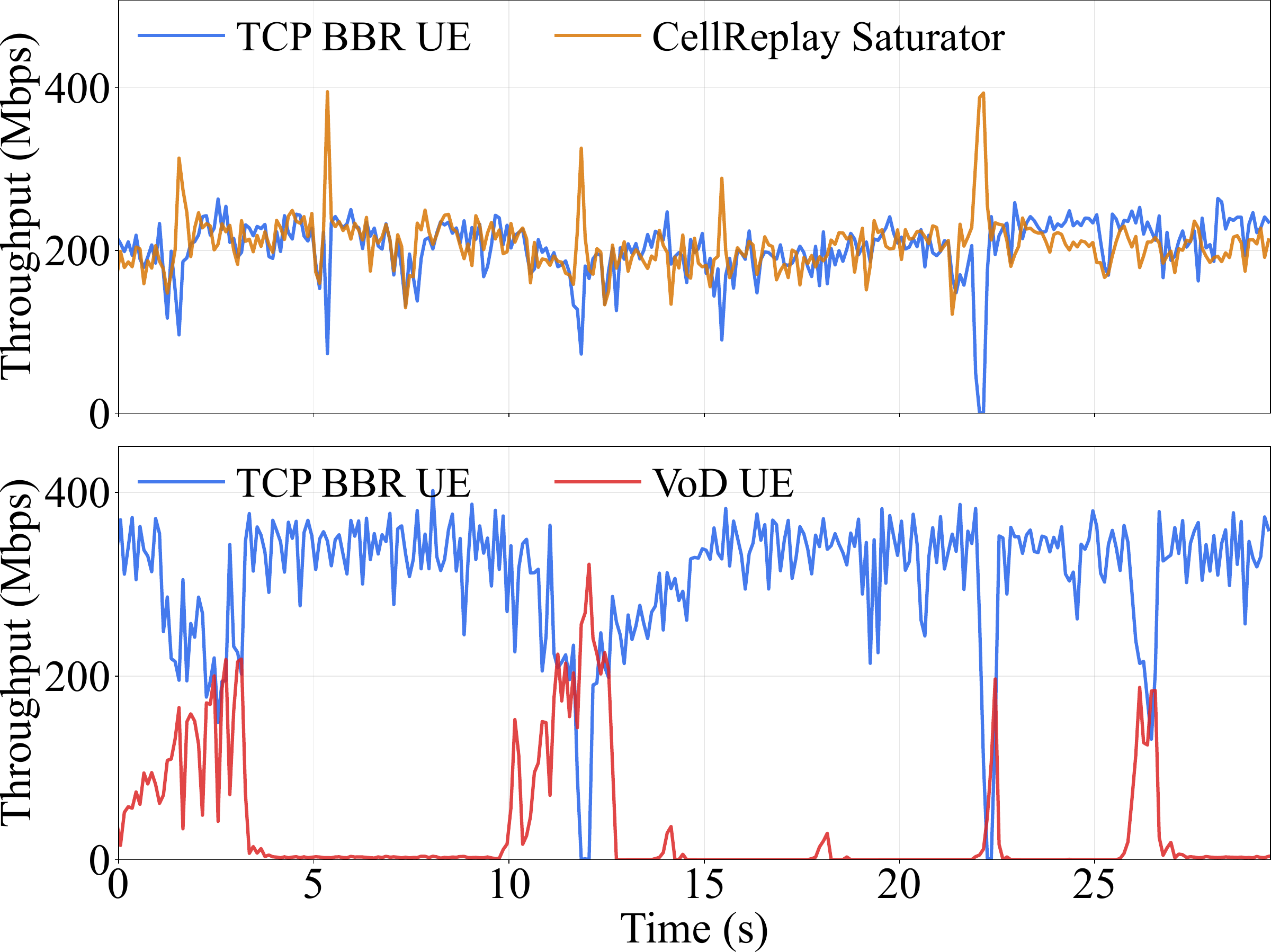}
        \caption{Throughput of VoD client coexisting with a BBR background flow.}
        \label{fig:contention_tput}
    \end{minipage}
    \hfill
    \begin{minipage}[b]{0.31\linewidth}          
       \centering
        \includegraphics[width=\linewidth]{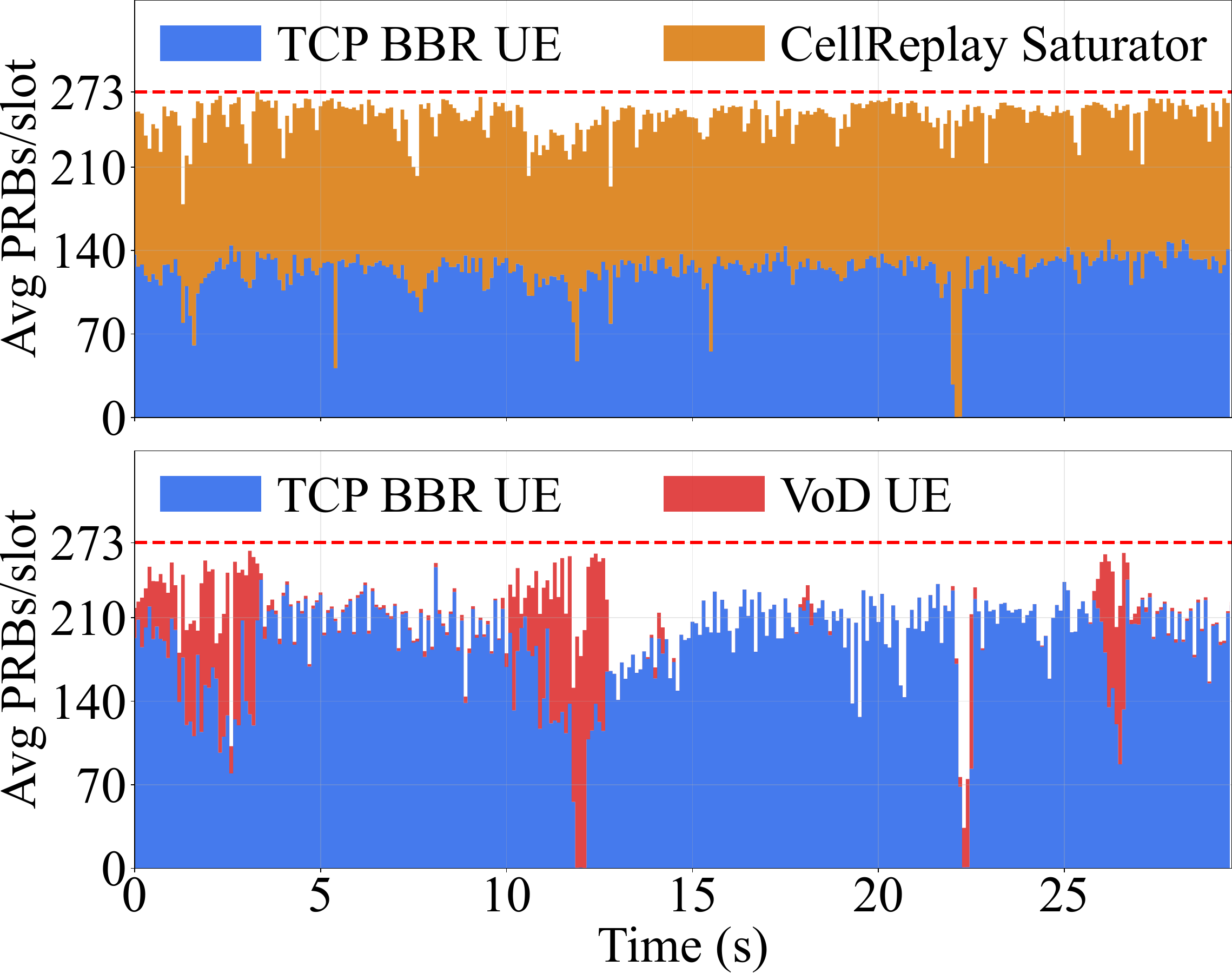}
        \caption{PRB allocation of VoD client coexisting with a BBR background flow.}
        \label{fig:contention_prb}   
    \end{minipage}
    \hfill
    \begin{minipage}[b]{0.33\linewidth}          
       \centering
        \includegraphics[width=\linewidth]{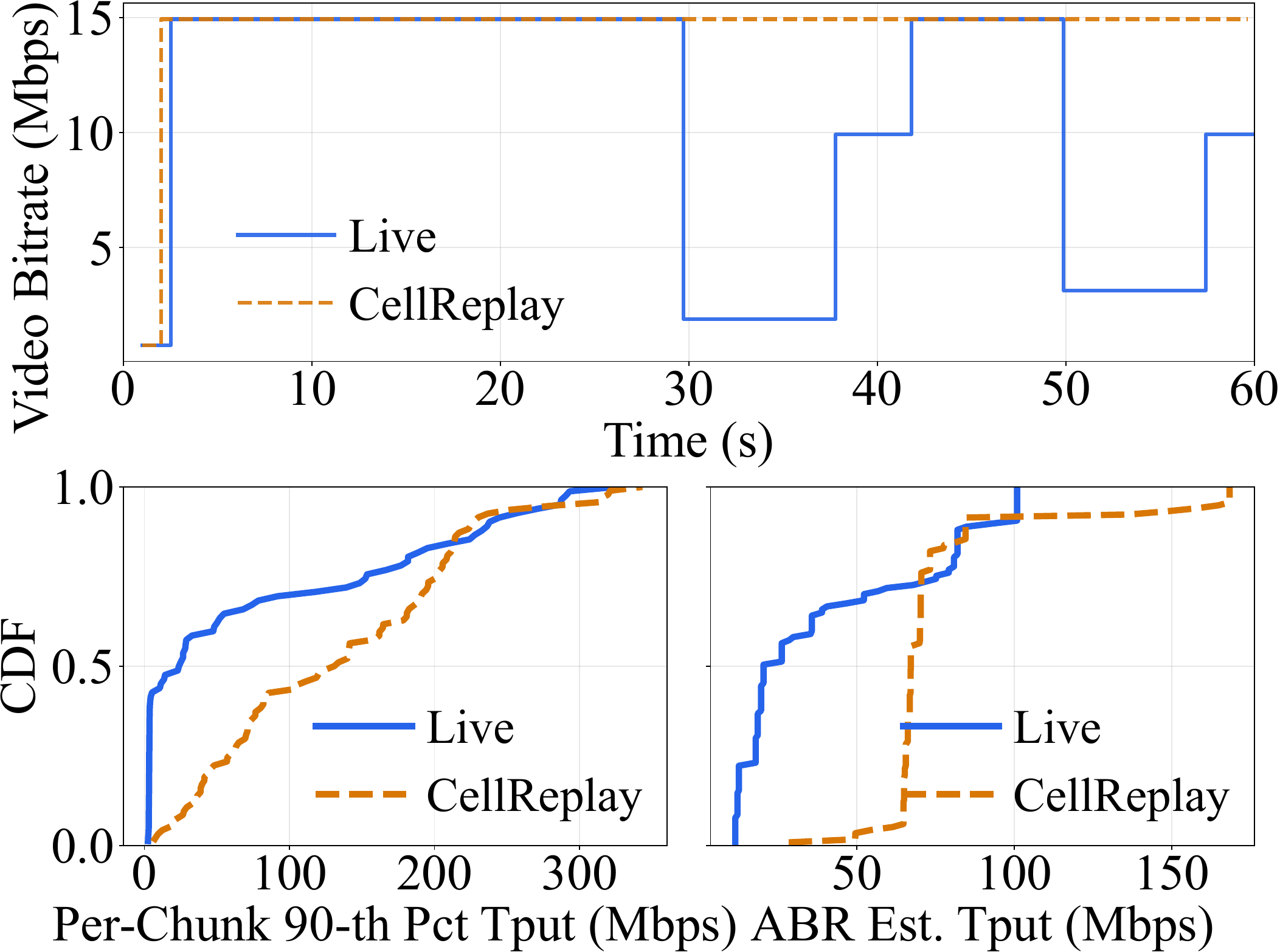}
        \caption{Live VoD compared with the CellReplay emulation.}
        \label{fig:contention_emulation}   
    \end{minipage}
    \vskip -0.15in
\end{figure*}

\section{Motivation}


Unlike a wired link that services queues at a relatively constant rate,
a commercial 5G base station acts as an intelligent, highly dynamic bottleneck.
To understand why evaluating network algorithms in such an environment is so challenging,
we must first examine the fundamental relationship between
the cellular network's scheduler and the applications it serves.

\subsection{Network-Application Coupling}
In a commercial cellular network, the base station is not a simple pipe; rather,
it actively and intelligently manages a shared pool of wireless resources.
As introduced in \S\ref{s:background:5g}, the base station assigns an independent packet buffer to each
connected UE continuously monitoring
its individual channel conditions (\textit{e.g.}, CQI).

\parahead{Scheduler adapts to the application.} 
The base station's scheduler acts dynamically upon these individual buffer and channel states.
When an application alters its sending behavior,
for instance, a TCP flow ramping up during slow start,
the corresponding per-UE RAN buffer quickly accumulates.
The scheduler observes this growing queue and, balancing complex,
vendor-specific fairness constraints with the traffic demand,
reacts by allocating more PRBs to this UE to drain the queue.
Conversely, if an application sends data slowly, its buffer remains shallow,
and the scheduler will dynamically reassign resources to other, 
more demanding background users.

\parahead{Application adapts to the network.} 
This dynamic resource allocation directly dictates the end-to-end performance 
perceived by the content server and UE,
\textit{i.e.}, the instantaneous throughput and packet delay.
Crucially, modern end-to-end applications and protocols,
such as ABR logic for video streaming or advanced congestion control algorithms,
operate by continuously measuring these exact performance metrics.
When the scheduler provides more resources, the application observes lower delay and higher throughput,
prompting it to increase sending rate or video resolution.

\parahead{Takeaway:}
This bidirectional interaction essentially forms a \textit{closed-loop feedback control} system 
between the network and the edge application: 
the network's scheduling behavior dictates the application's performance,
and the application's resulting traffic pattern acts as a feedback
signal that fundamentally reshapes the network's future scheduling decisions.

\subsection{Limits of Record-and-Replay}
The closed-loop feedback control between network protocols and the RAN
scheduler creates fundamental issues for traditional trace-driven or record-and-replay emulators,
which are currently the state of the art for end-to-end protocol evaluation.
Tools such as Mahimahi 
\cite{netravali_mahimahi_2015} and CellReplay 
\cite{sentosaCellReplayAccurateRecordandreplay2025}
rely on an inherently \textit{aggressive} probing strategy to capture network characteristics.
To discover the absolute upper bound of the network's capacity (packet delivery opportunities, PDO),
their \textit{saturator} module floods the connection,
pushing as much data as possible to ensure the RAN buffer is continuously full during the recording phase.

However, for the RAN,
this aggressiveness creates a measurement inaccuracy: the act of
saturating the network fundamentally alters the network's behavior.
Because of the tight network-application coupling described above,
the scheduler observes the saturator flow's artificially large buffer and reacts
by highly prioritizing this flow to clear the backlog.

Furthermore, since the interaction between an application and the 
RAN is stateful and unrepeatable, 
a network trace is merely a snapshot of the RAN scheduler's
reaction to one specific application's traffic pattern at one specific moment.
When record-and-replay tools attempt to evaluate a completely
different target application using this frozen trace,
they make a \textit{fundamentally flawed assumption}: that the network will blindly provide
exactly the same resource allocation regardless of the new application's actual behavior.

Summarizing, since they treat a dynamic, stateful interaction as a static log,
these emulators force target applications to experience a pre-recorded,
highly-inflated transmission schedule that the live 5G network would never have actually assigned to a normal,
less aggressive workload.


%

\subsubsection{Empirical Evidence of Emulation Failure}
\label{s:motivation:failure}
To empirically demonstrate how this measurement inaccuracy
manifests in practice and distorts algorithm evaluation,
we contrast the performance of two distinct applications,
\textit{i.e.}, Video on Demand (VoD) and the Copa congestion control algorithm,
in a live commercial 5G network versus a state-of-the-art record-and-replay emulator: 
CellReplay~\cite{sentosaCellReplayAccurateRecordandreplay2025}.

\parahead{Case Study 1: Failure to emulate sharing dynamics.} 
Because real-world applications rarely operate in isolation,
a faithful emulator must accurately replicate how network resources are shared among competing users.
To demonstrate how current record-and-replay tools fail to capture these phenomena,
we evaluate how a VoD application performs when coexisting with a single, aggressive 
BBR background flow.
To cleanly isolate this cross-user contention,
we conduct this experiment late at night with 
quiet background traffic.

During the network trace recording phase,
CellReplay runs its saturator tool alongside the BBR flow.
Since the saturator sends as fast as possible, 
it competes evenly with the aggressive BBR flow.
Consequently, the RAN scheduler splits capacity roughly
equally between the two (\cref{fig:contention_tput} \textit{top} and 
\cref{fig:contention_prb} \textit{top}).
Since the emulator assumes this recorded capacity would 
be available to any future sender, during the subsequent replay phase,
it replays this artificially-inflated \textit{equal split} to the VoD client.
The VoD's internal ABR logic then significantly overestimates
available network capacity (\cref{fig:contention_emulation} \textit{bottom}).
Driven by this false estimate,
the VoD application increases and maintains 4K resolution (15~Mbps,
\cref{fig:contention_emulation} \textit{top}).

In the real 5G network, however, the dynamic is entirely different.
A VoD application is inherently polite and bursty:
downloading a chunk and then idling,
while BBR is persistently greedy.
The RAN observes the VoD's mostly empty buffer (\cref{fig:contention_tput} \textit{bottom}) 
alongside BBR's deep queue, and rightfully allocates
the vast majority of the bandwidth to BBR (\cref{fig:contention_prb} \textit{bottom}).
Crucially, this creates a constrained feedback loop. As the VoD client receives limited per-chunk throughput,
its internal ABR logic correctly estimates a highly constrained network capacity (\cref{fig:contention_emulation} \textit{bottom}). 
Reacting to this,
the VoD client conservatively selects a much lower video bitrate (\cref{fig:contention_emulation} \textit{top}).
Because the VoD client remains in this low-bitrate, polite state, its buffer remains consistently shallow,
continuously reinforcing the base station's decision to favor the greedy BBR flow over it.

\parahead{Takeaway:} Ultimately, this experiment exposes the 
core fallacy of trace-driven emulation: treating capacity as a
static environmental constant rather than a dynamic outcome of the application-network coupling.
By freezing the scheduler's state, static traces sever the continuous feedback loop,
fundamentally invalidating the evaluation of any new workload.

\begin{figure}
    \centering
    \includegraphics[width=0.5\linewidth]{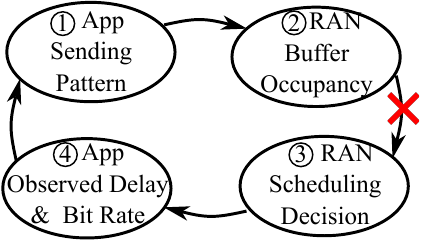}
    \caption{Causality graph of RAN and applications: record-and-replay breaks RAN scheduling decision causality.}
    \label{fig:causality_graph}
\end{figure}

\parahead{Case Study 2: Failing to capture behavior causality.} 
The record-and-replay approach can't capture the RAN's dynamic adjustment of its scheduling decisions, 
exposing only fixed behavior based on the trace and breaking the loop of
\cref{fig:causality_graph}.

\begin{figure}
    \centering
    \includegraphics[width=\linewidth]{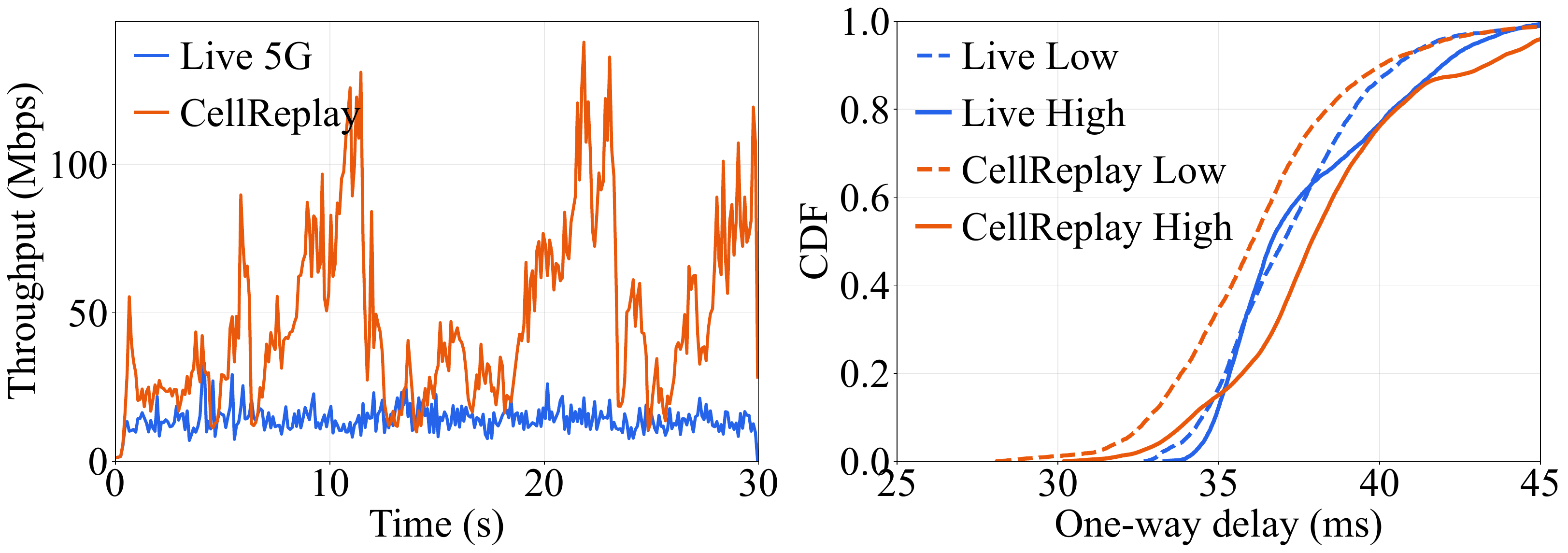}
    \caption{\textit{Left}: Copa's achieved throughput in Live 5G and CellReplay. \textit{Right}: Copa's observed one-way delay for low (< 10-th pct.) and high (> 90-th pct.) packet burst sizes.}
    \label{fig:causality_copa}
    \vskip -0.15in
\end{figure}

\begin{figure}
    \centering
    \includegraphics[width=\linewidth]{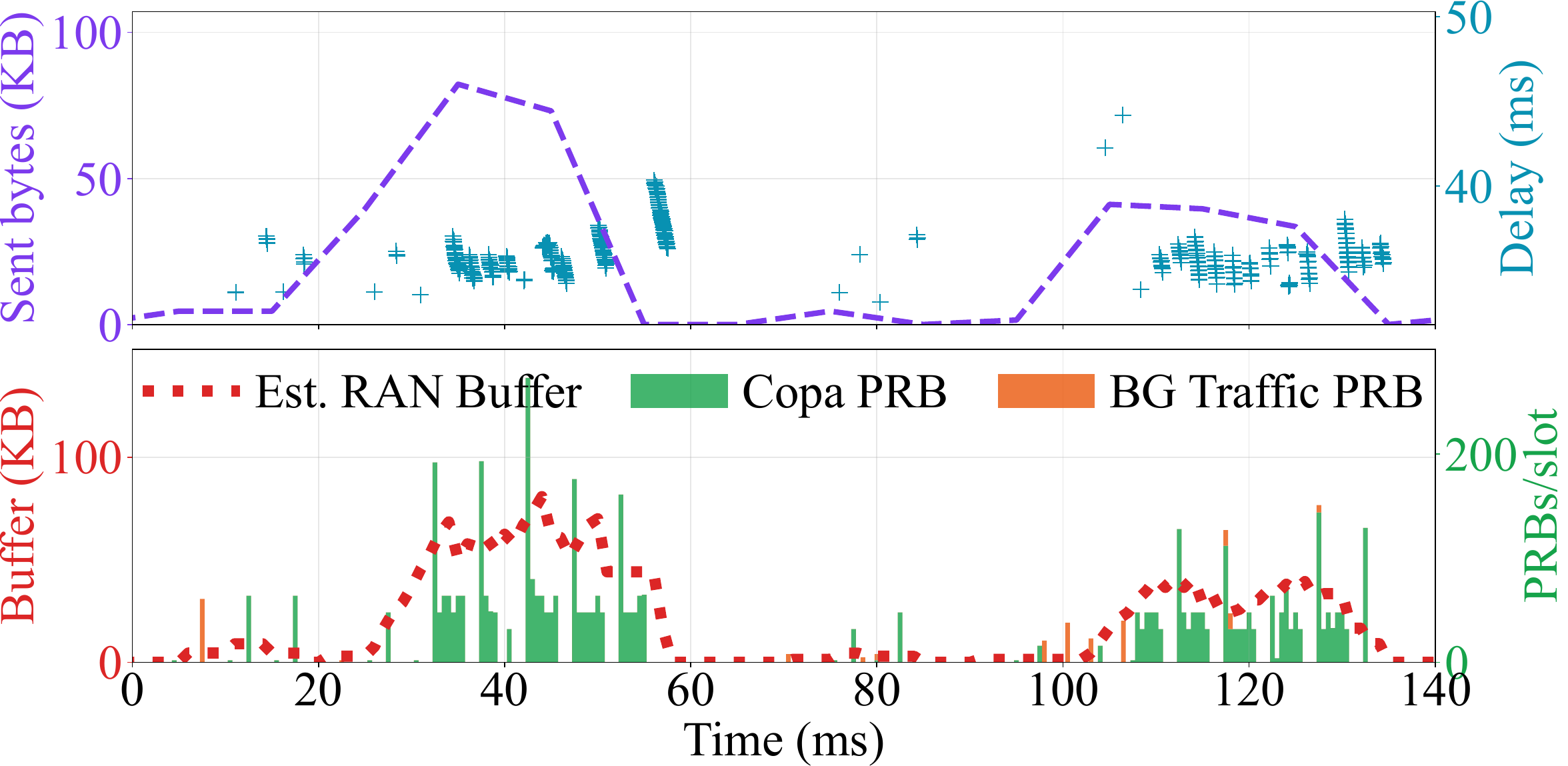}
    \caption{Copa's behavior of two sending data bursts in the live 5G network. PRB measurement collected with NR-Scope.}
    \label{fig:causality_copa_bursts}
    \vskip -0.15in
\end{figure}

Copa\cite{arun_copa_2018} in a production T-Mobile
network (during steady state) almost fully drains the path bottleneck's 
queue and then refills it every 2\sfrac{1}{2} RTT cycles \cite{arun_copa_2018}, setting
its send rate to $R_s = 1/d_q$, where $d_q$ is the estimated mean per-packet 
queuing delay.
We configure a Copa receiver in a local laptop tethered to a UE,
and the sender in a Google Cloud Platform (GCP) server.
We run Copa for 30 seconds, separately 
collect CellReplay traces with its saturator, and then replay Copa in CellReplay in a less busy network.
The two ends are time synchronized to the GPS signal and 
during the experiment, we collect RAN scheduling telemetry
with NR-Scope and a network trace with \texttt{tcpdump}. 
After data collection, we estimate the UE's RAN buffer by shifting the cumulative server egress bytes by one-way delay and subtracting the cumulative client ingress bytes from the result.
Even though the RAN can provide up to 270~Mbps for a single user, 
Copa achieves a throughput of only 15~Mbps (\cref{fig:causality_copa} \textit{left}).
\cref{fig:causality_copa_bursts} shows a more detailed behavior in Copa and the RAN, where the RAN reacts to Copa's different buffer sizes \circled{2} and allocates fewer resources when the buffer occupancy is lower \circled{3} (100 to 120 ms in upper and lower \cref{fig:causality_copa_bursts}) compared to the higher buffer occupancy (20 to 60 ms in upper and lower \cref{fig:causality_copa_bursts}).
Thus, the Copa endpoints observe similar delays in both scenarios (upper in \cref{fig:causality_copa_bursts}), which appears to be a phantom "standing" queue to Copa \circled{4}, thus the sender doesn't increase its sending rate \circled{1}.
\cref{fig:causality_copa} (right) shows the one-way delay distribution of low (< 10-pct.) and high (> 90-pct.) burst sizes of Copa in live 5G network and CellReplay emulation.
The live 5G network results in a similar one-way delay distribution for the low and high bursts with EMD 0.54ms, which is caused by the scheduler's work-load dependent behavior and retransmission (\circled{2} $\rightarrow$ \circled{3}).
CellReplay's emulation shows clear distinction between the low and high bursts with EMD 2.01ms, meaning the distribution shifts about 3.7$\times$ more between low and high than the live 5G link.
As a result, Copa achieves $\sim4\times$ overestimated throughput in CellReplay.


\parahead{Takeaway:}
This experiment exposes a critical flaw of record-and-replay: 
it severs the causal feedback loop between the sender 
and the RAN's dynamic resource allocation. 


\subsection{Limits of Full-stack Emulation}
To preserve the crucial closed-loop interaction between the application and the network,
an alternative approach is whole-system, full-stack simulation or emulation,
utilizing platforms such as ns-3~\cite{nsnamNs3}, srsRAN~\cite{system_srsran_2023}, 
and OpenAirInterface (OAI)~\cite{OaiOpenairinterface5GGitLab2026}.
Unlike record-and-replay methods,
these systems model the complete cellular protocol stack—from the physical layer up to
the core network—ensuring that network allocations dynamically react to the application's sending behavior.
However, this approach falls short in two critical dimensions.

\begin{figure}
    \centering
    \subfigure[RTT results.]{
    \includegraphics[width=0.49\linewidth]{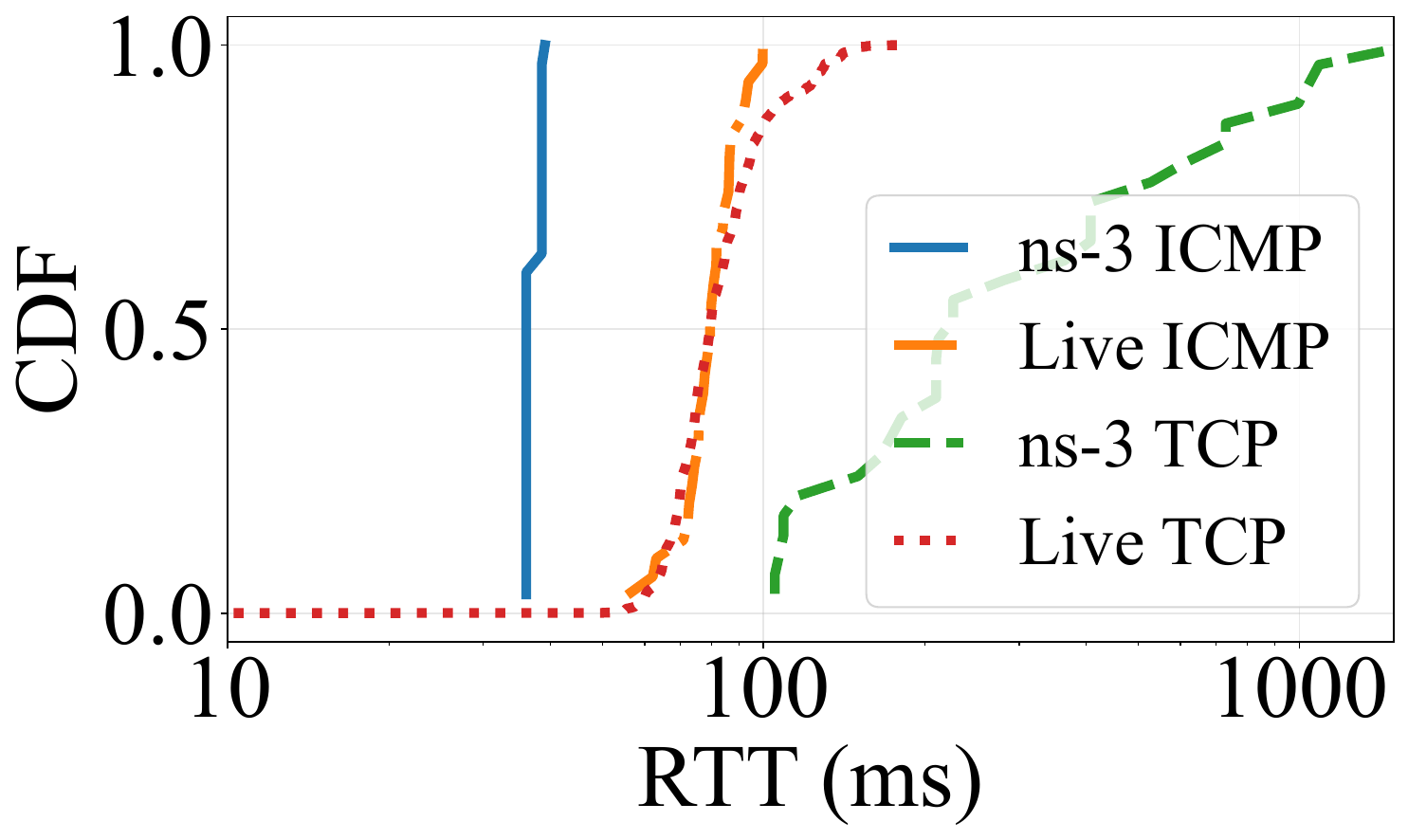}\label{fig:rtt_emu}}
    \hfill
    \subfigure[Throughput results.]
    {\includegraphics[width=0.49\linewidth]{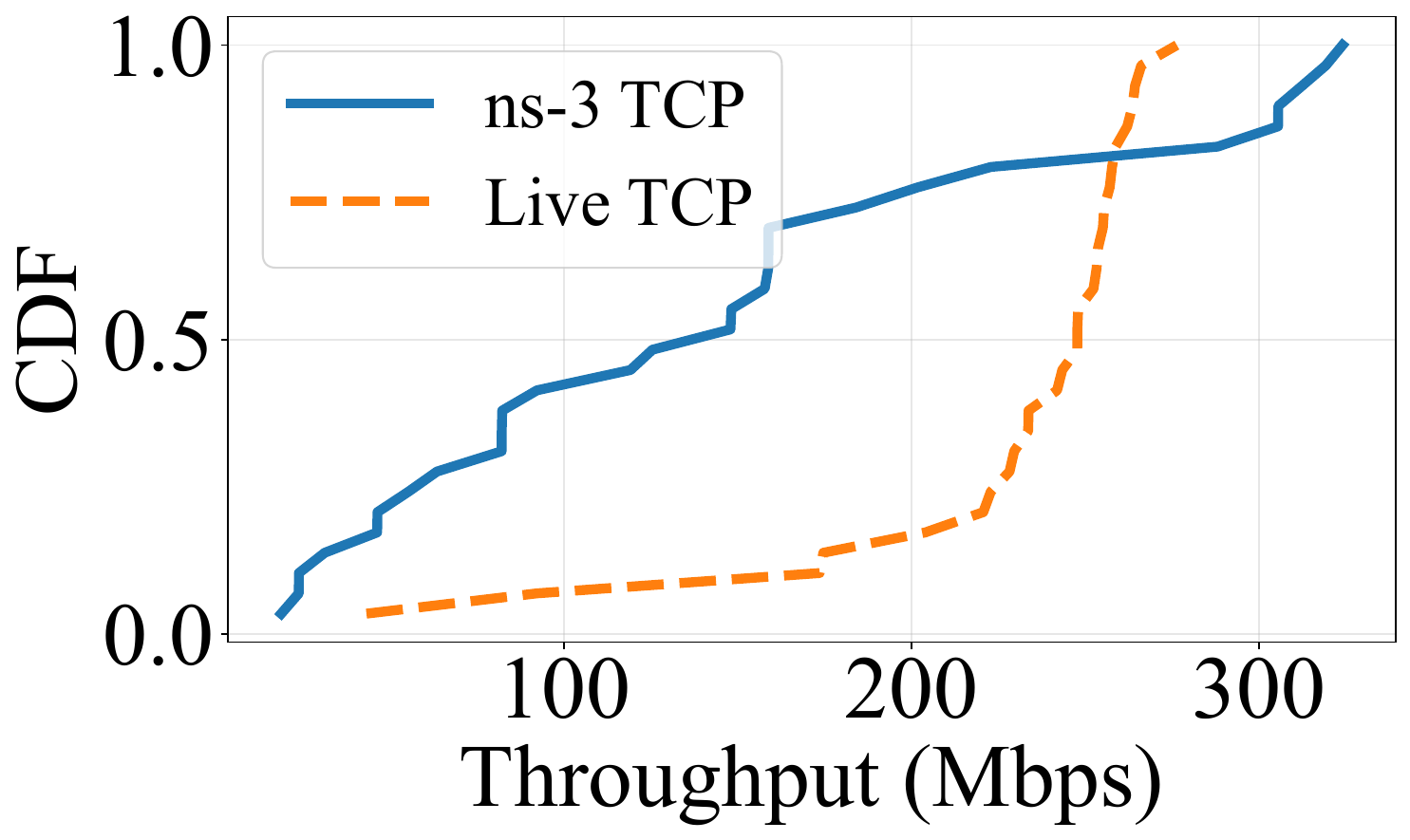}\label{fig:tput_emu}}
    \caption{RTT and throughput for \textbf{ns-3 simulation} and production 5G 
    with the same PHY configuration.}
    \label{fig:ns-3_simulation} 
    \vskip -0.15in
\end{figure}

First, they suffer from a severe realism gap at the scheduler.
Commercial 5G base stations rely on highly complex,
proprietary scheduling logic designed to balance vendor-specific KPIs
and user fairness in highly dynamic radio environments.
Open-source cellular protocol stacks, conversely, are forced to implement simplistic,
textbook algorithms, such as basic round-robin~\cite{arpaci-dusseauOperatingSystemsThree2018} or 
proportional fair~\cite{kushnerConvergenceProportionalfairSharing2004}.
To demonstrate this disparity,
we ran an experiment with ICMP and TCP traffic in a live T-Mobile 5G network,
and compared the results against an ns-3 simulation
configured with identical physical layer parameters 
(bandwidth, subcarrier spacing, MCS, \textit{etc.}).
As shown in \cref{fig:ns-3_simulation}, even with the full protocol stack implemented,
ns-3 fundamentally fails to capture the latency and throughput dynamics of a real commercial network.
Specifically, the commercial 5G network achieves significantly higher TCP throughput,
demonstrating that the simplistic schedulers in open-source simulators lack
the advanced optimizations and efficiency present in production networks.

Second, full-stack implementations introduce prohibitive computational overhead 
and software complexity.
Theoretically, one could achieve faithful emulation by implementing 
a highly accurate, commercial-grade scheduler within these platforms.
However, maintaining the entire protocol stack,
including signaling, encoding, and low-level physical layer processing,
is largely extraneous to evaluating end-to-end application performance.
These bloated architectures make deployment notoriously difficult 
and incur immense computational overhead.
For instance, running a basic OAI setup can spawn
over a thousand system threads, making lightweight, scalable,
and highly concurrent emulation impossible.

\subsection{A Scheduler-Driven Digital Twin}
The limitations discussed above reveal a clear \textit{design mandate}: 
a faithful, lightweight cellular emulator must 
perfectly replicate the commercial scheduler's decision-making
process without implementing the entire, bloated protocol stack.

We propose a new paradigm: \textit{scheduler-driven emulation}.
Because the MAC scheduler's dynamic resource allocation and basic link-layer mechanisms 
(\textit{e.g.}, HARQ retransmissions) are the root causes of end-to-end throughput volatility and delay jitter,
we only need to accurately model these specific components.
By treating the proprietary commercial scheduler as a black-box, 
we can map the RAN's recent per-UE states,
\textit{e.g.}, buffer occupancy, channel quality measurement, PRB allocation, and selected bit rate,
directly to its corresponding scheduling decisions,
\textit{e.g.}, PRB allocation and bit rate selection in the near future:
\begin{equation}
    f_1: \{(\bm{b}, \bm{c}, \bm{p}, \bm{m}, \bm{s})[i-d: i]\} \rightarrow \{(\bm{p}, \bm{m})[i+1:i+w]\}, \label{eq:scheduler_mapping}
\end{equation}
where the RAN scheduler allocates resources for each UE for the next $w$ slots,
based on high-resolution measurements of the past $d$ slots' RAN states, as detailed in~\cref{tab:variables}.
In this way, we employ Machine Learning to understand and 
replicate the exact behavior of this black box without actually 
needing to open it or decode its proprietary algorithms.
Successfully capturing this behavioral mapping is the critical step
that enables us to build a highly accurate,
scheduler-focused digital twin.
By coupling this learned scheduler with essential,
performance-critical protocol operations (which we detail in \S\ref{s:impl}),
we can entirely bypass the immense computational overhead and complexity of full-stack simulators.
\begin{table}[]
\small
\begin{tabularx}{\linewidth}{@{}cl}
\toprule
\textbf{Variable}& \textbf{Meaning}\\\midrule
$b_k[i]$ ($\bm{b}[i]$)    & Buffer occupancy for user $k$ (all users).\\
$c_k[i]$ ($\bm{c}[i]$)    & Channel quality indicator (CQI) for user $k$ (all users).\\ 
$p_k[i]$ ($\bm{p}[i]$)    & Allocated PRB for user $k$ (all users).\\ 
$m_k[i]$ ($\bm{m}[i]$)    & Selected MCS index for user $k$ (all users).\\ 
$s_k[i]$ ($\bm{s}[i]$)    & TBS of user $k$ (all users).\\ 
$a_k[i]$                  & RAN ingress data size for user $k$.\\
\bottomrule
\end{tabularx}
\caption{Variables used in \sysnames{} design development: variable $i$ 
indexes 5G time slots.} 
\label{tab:variables}
\vskip -0.15in
\end{table}

\parahead{Challenges and Roadmap.} While framing the problem as a supervised learning task is intuitive,
compiling the requisite high-resolution, multi-user training data constitutes the most formidable challenge.
We can partially observe the network's behavior using state-of-the-art telemetry tools: NG-Scope \cite{xie_ng-scope_2022}
 and NR-Scope \cite{wan_nr-scope_2024} allow us to passively sniff downlink control information,
capturing the scheduler's final decisions (allocated PRBs, $\bm{p}[i]$, and MCS, $\bm{m}[i]$) for all users in the cell.
Additionally, we can instrument our own controlled endpoints with tools
like CellNinja~\cite{liuSeeingFogEmpowering2025} to record their local CQI ($c_k[i]$).

However, a vast amount of critical input data remains strictly hidden inside the commercial base station.
Most notably, the base station's internal state,
specifically the instantaneous buffer occupancy ($\bm{b}[i]$) for every user,
is completely opaque to the outside world.
Furthermore, a commercial cell is heavily shared with uncontrolled background users whose
individual channel states and incoming traffic demands are entirely invisible.
Without full visibility into these hidden queues and background demands,
it is impossible for the ML model to learn the true cross-user contention dynamics that drive the scheduler.
Overcoming this severe missing data problem,
by estimating our own buffers and mathematically
reconstructing the hidden background traffic,
is the key to unlocking our scheduler-driven twin,
which we detail in \S\ref{s:design}.

\section{Design}\label{s:design}

With the problem set up, we now elaborate our design.
We start with the collection of the required data from the 
RAN (\S\ref{s:design:data_collection}), and then discuss the ML model 
designs (\S\ref{s:design:traffic_reconstructor} and \ref{s:design:scheduling_learner}).

\subsection{Data Collection and Processing} \label{s:design:data_collection}

ML models require massive volumes of data to learn effectively. 
Our high-resolution, time-synchronized measurement tools 
observe the RAN from the outside, simultaneously capturing data 
at the application, device, and radio layers.

\parahead{Capturing the observable.}
We capture telemetry data:
\begin{itemize}
    \item The network's view: We deploy a software-defined radio running NR-Scope to passively eavesdrop on the 5G air interface. 
    This grants us a global view of the base station's per-slot resource allocation ($\bm{p}[i]$ and $\bm{m}[i]$).
    \item The device's view: For our controlled endpoints (tethered smartphones), we run CellNinja \cite{liuSeeingFogEmpowering2025} to extract diagnostic modem logs. 
    This reveals the CQI ($c_{1:k}[i]$) our devices (1 to $k$) report back to the tower—the crucial environmental variable driving the scheduler's MCS selection.
\end{itemize}
These two sources of data are already aligned by the 5G network's
$500\ \mu s$ slot time base.
Separate 5G base stations are also synchronized to GPS time
for cross-RAN mobility, such as handover between 
base stations\cite{pultarova_5g_2026}.

\parahead{Inferring the invisible.}
With these RAN telemetry data collected, we still need to acquire the 
per-slot buffer occupancy ($\bm{b}[i]$), which is not directly collectible 
from any source other than the base station itself.
To crack this problem, we model packet flow timing and infer buffer occupancy.

\begin{figure}
    \centering
    \includegraphics[width=\linewidth]{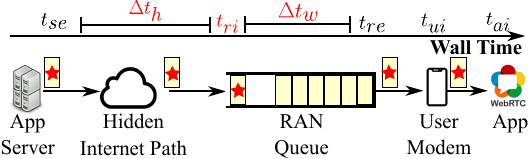}
    \caption{The packet's journey from the server to the client over 5G network, we can't directly measure the {\color{red} red} values.}
    \label{fig:pkt_timing}
    \vskip -0.15in
\end{figure}

First assuming that all the components are synchronized with 
a universal clock (wall time), we follow the lifetime
of the packet marked with {\color{red} $\bigstar$} 
in \cref{fig:pkt_timing}. 
At the content server, we measure the \underline{s}erver \underline{e}gress time 
$t_{se}$ with \texttt{tcpdump}.
Then the packet goes through the \underline{h}idden internet 
path with delay of $\Delta t_{h}$. 
For model training purposes, we choose and verify this path to be 
higher speed and more stable in one-way delay compared to the 
RAN, so that $\Delta t_{h}$ remains constant to a single-millisecond variance 
during the data collection session (approximately two minutes).
Then the packet \underline{w}aits in the RAN-UE buffer for $\Delta t_{w}$, 
comprising the queuing delay, scheduling delay and non-congestive 
delay.\footnote{\textit{Queuing delay}: time to reach the queue head. 
\textit{Scheduling delay}: time waited at the head of the queue, as the 
RAN is scheduling other users. \textit{Non-congestive delay}: when
the first transmission fails, the RAN retransmits the data, 
creating non-congestive delay.}
The \underline{R}AN \underline{e}gresses the packet at $t_{re}$, 
which is exactly the same as the \underline{i}ngress 
time of the packet into \underline{u}ser modem $t_{ui} = t_{re}$.
We measure $t_{ui}$ and $t_{re}$ with the telemetry tools (NR-Scope and CellNinja).
Then the packet exits the modem to the kernel, 
and then the \underline{a}pplication, at which point 
\texttt{tcpdump} records time $t_{ai}$.
The user's modem holds the packets, waiting for retransmissions, 
to guarantee in-order delivery to the kernel, creating the time 
differences between $t_{ui}$ and $t_{ai}$.
As the target, we want to estimate $\Delta t_h$ and then 
$t_{ri}$ of each packet so that we can jointly estimate the buffer 
size at any time with the packet egress time $t_{re}$.
In this scenario, the internet path transmission time ($\Delta t_h$) is 
relatively fixed and stable, while the waiting time ($\Delta t_w$) in 
the RAN is highly volatile.
We use the global minimum one-way delay, $\min(t_{re} - t_{se})$ 
within each measurement session, as the estimate of $\Delta t_h$, 
corresponding to when the RAN queue is empty and the 
transmission is successful in one pass.

\begin{figure}
    \centering
    \includegraphics[width=\linewidth]{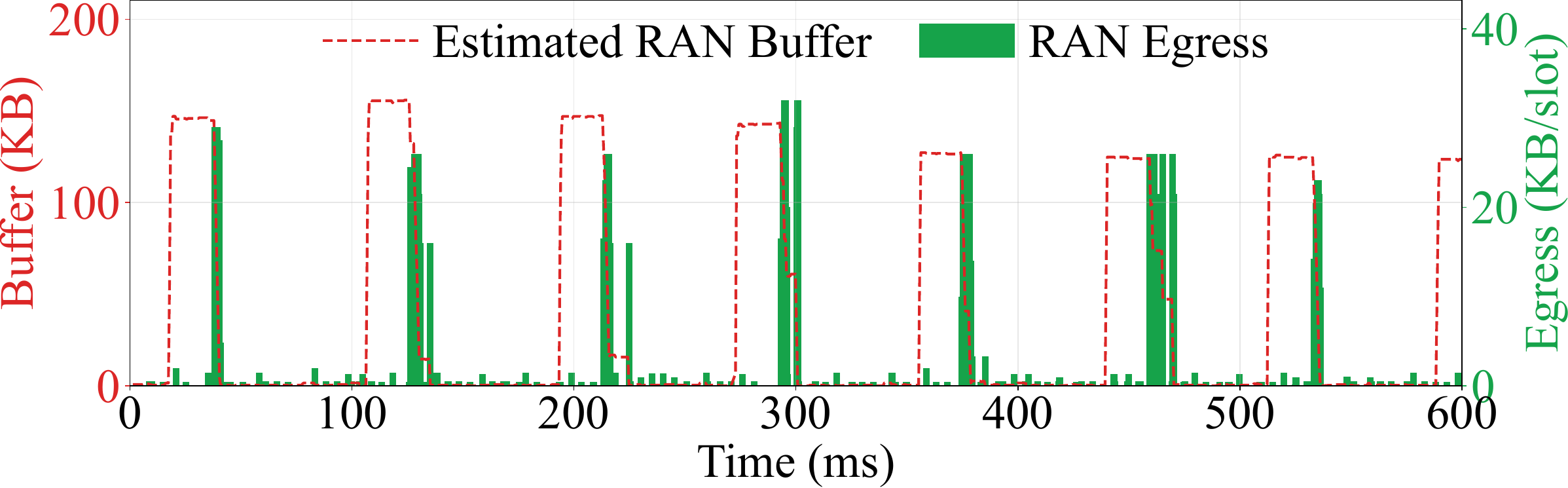}
    \caption{Buffer estimates of example cloud gaming traffic.}
    \label{fig:buffer_example}
    \vskip -0.25in
\end{figure}

Returning to reality, where there is no perfect global clock, 
UEs are time synchronized to the base station in 
slots ($t_{re}$ and $t_{ui}$ observed by the telemetry tools), 
which is ultimately synchronized to the GPS signal.
To synchronize the user's kernel time ($t_{ai}$) with the RAN telemetry 
data, we set up a Raspberry Pi to receive the GPS signal and distribute
its time through a local L2 switch to the tethered laptops.
As for the remote server, we use the Google Cloud Platform's internal
NTP server, reported to be synchronized with their 
atomic clock with a GPS receiver \cite{ConfigureNTPCompute,PublicNTP}. 
We don't use NTP over the cellular link because the user's wireless link 
creates volatile and asymmetric RTTs, skewing its calculations.
GPS 
approaches distributed synchronization with a $50\ \mu s$ theoretical 
time offset---well within the $500\ \mu s$ slot time.
However, tiny time drifts happen during data collection, creating 
delayed or advanced packet egress, but
since these drifts happen rarely and slowly, \sysname{} 
tolerates them.
With the timing issue resolved, we estimate buffer occupancy
($b_{1:k}[i]$) of our controlled $k$ users at any time $i$ by 
comparing the estimated $t_{ri}$ and measured $t_{re}$ of every packet.
\cref{fig:buffer_example} shows an example of the 
estimate for cloud gaming streaming 15 FPS video.  

\parahead{Discussion.}
In this data collection pipeline, the \textit{egress data} are accurately extracted and from the Qualcomm modem through CellNinja \cite{liuSeeingFogEmpowering2025} (for the controlled user), and from the raw wireless signal through NR-Scope \cite{wan_nr-scope_2024} (for uncontrolled user).
NR-Scope achieves good data collection accuracy (below 4~\% missing rate), and we place the SDR at a location with the best signal quality in our lab.
For the \textit{ingress data} estimation, GPS provides accurate timing, and we select a close cloud server to reduce internet path variation.
Given our goal is to estimate the RAN queue occupancy for training, instead of generalizability, this selection serves our goal properly.
Combining the accurate ingress and egress data, we can estimate the queue occupancy with an acceptable accuracy, and the end-to-end evaluation results confirm that.

\parahead{What's still missing.} With this setup, we can collect the scheduling results for all the users in the network ($\bm{p}[i]$ and $\bm{m}[i]$), and the RAN-UE states for our controlled users ($c_{1:k}[i]$ and $b_{1:k}[i]$). 
To learn the all-around behavior of the RAN scheduler, we still lack the background users' 
state data ($c_{k+1:K}[i]$ and $b_{k+1:K}[i]$).
However, since we don't control the remote server of the background users, we can't estimate their buffer occupancy the same way as our controlled users.
For this, we train an ML model with 4 controlled users' data (\S\ref{s:impl}) to recover the buffering pattern for the background users.

\subsection{Traffic Reconstructor} \label{s:design:traffic_reconstructor}
While our controlled testbed successfully captures the complete state of our own devices, it leaves a critical blind spot: the background users. 
Because we do not control the remote servers generating their traffic, we cannot directly measure their incoming packet patterns or reliably estimate their buffer occupancies. 
We can, however, eavesdrop on the base station's final scheduling decisions for all users across the network via the telemetry data. 
To complete our holistic view of the cell's dynamics, we introduce \trafficmodel{}, an ML model designed to bridge this gap by solving the inverse problem.

\parahead{Working backward from telemetry.}
The \trafficmodel{} acts as a decoder, working backward from the observable scheduling results (PRB 
allocation, MCS index and TBS) to recover the inputs that led to those results:
the user's internal CQI ($c_k[i]$) and the base station's ingress data size ($a_k[i]$), 
from which we subtract the RAN's egress bytes (TBS, $s_k[i]$) to precisely reconstruct 
buffer occupancy over time.

A natural question arises: why not train the model to predict the buffer level $b_k[i]$ directly?
The fundamental issue is that absolute buffer level at any given slot is a cumulative 
quantity---it is the net difference between globally accumulated ingress and egress bytes
over time ($b_k[n] = \sum_{i=0}^{i=n} (a_{k}[i] - s_k[i])$), and so time-localized
ML models struggle to infer this value accurately.
For instance, when a steady state flow's incoming 
rate perfectly matches its egress rate, buffer size hovers constant, rendering 
it effectively invisible to a stateless model.

\begin{figure}
    \centering
    \includegraphics[width=\linewidth]{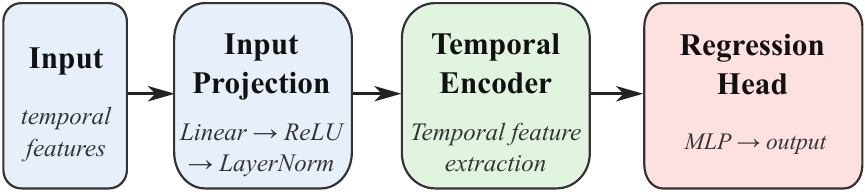}
    \caption{\trafficmodel{} model structure.}
    \label{fig:traffic_reconstructor}
    \vskip -0.15in
\end{figure}

\parahead{Mathematical formulation and architecture.}
By shifting the target to the instantaneous incoming packet size and CQI, we frame a much more stable learning objective.
Mathematically, the model achieves the following inverse mapping:
\begin{equation}
    f_2: \{(p_k, m_k, s_k, p_{\{1:K\}/k})[i-h: i+h]\} \rightarrow \{(a_k, c_k)[i]\},
\end{equation}
where $p_{\{1:K\}/k}[i]$ is the sum total number of PRBs allocated for all other users except the target user $k$.
Because this inverse mapping occurs offline after the data collection, the model has the luxury of examining a window of both past and future scheduling decisions to infer the precise ingress data size and CQI.
Furthermore, since individual user channel status and ingress patterns are generally independent, we apply this model on a per-user basis.
We treat each controlled flow individually and ask the model to recover the flow's time-series ingress data from the egress patterns.
We split the data into training, validation, and test set, and select the model that has best accuracy and generalizability.
Architecturally, we process the features through the input projection layers, utilize a temporal encoder to extract complex time-series features, and finalize the prediction through a regression head (\cref{fig:traffic_reconstructor}).

\parahead{Completing the data.}
In summary, by successfully solving this inverse problem, the \trafficmodel{} lifts the veil on background network dynamics. 
With the complete, multi-user RAN state now reconstructed—covering both our controlled endpoints and the surrounding uncontrolled users—we finally have the necessary data foundation. 
The next crucial step is modeling how the base station actually acts upon this complex state. 
To achieve this, we introduce \schedulermodel{}, the module responsible for translating these high-dimensional UE states into concrete, millisecond-level resource allocations.

\subsection{Neural Scheduler} \label{s:design:scheduling_learner}
With the complete, multi-user RAN-UE states reconstructed, the final problem is replicating the 5G network's core intelligence. 
The \schedulermodel{} serves as this core decision-making module.
As formalized earlier in \cref{eq:scheduler_mapping}, its objective is to learn the mapping ($f_1$) from historical network states to future resource allocations. 
By observing the past $d$ slots—user buffer occupancies, CQIs, and the scheduler's recent 
history (PRB, MCS, and TBS)—the model gains the crucial context needed to infer
both immediate workload demands and the network's underlying fairness constraints.

\begin{figure}
    \centering
    \includegraphics[width=\linewidth]{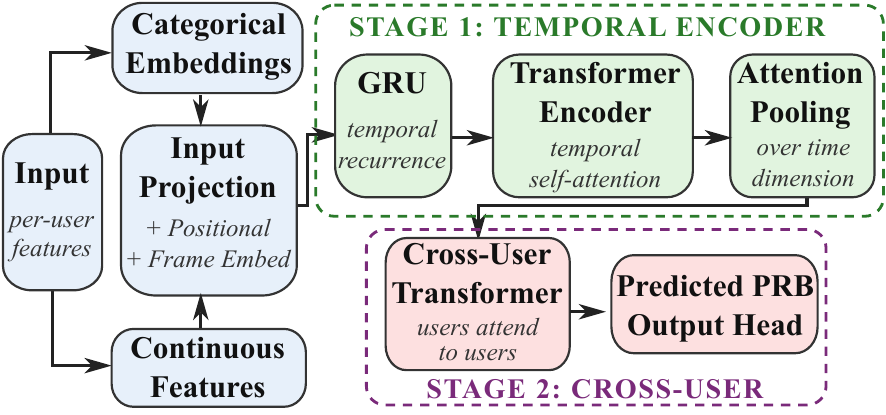}\label{fig:prb_predictor}
    \caption{\schedulermodel{} PRB model's architecture.}
    \label{fig:neural_scheduler_architecture} 
    \vskip -0.15in
\end{figure}

\parahead{Model architecture.}
Instead of forcing a single, monolithic model to untangle this massive feature space, we design
the \schedulermodel{} to reflect the dual nature of cellular resource management.
Cellular networks must balance two distinct realities: wireless channel 
quality is inherently individual, while spectrum resource allocation 
is highly competitive. 
Therefore, we divide the predictive workload into two specialized, lightweight neural networks:

\subparahead{PRB model:} This model is tasked with PRB allocation---since PRBs are a limited resource, the RAN decides how to divide them among multiple competing users. To capture this, the model needs to understand both a single user's history and how all users relate to each other at the current moment. We first use a Gated Recurrent Unit \cite{choLearningPhraseRepresentations2014} to extract each user's historical time-series features. Then, we pass these into transformer encoder layers \cite{vaswaniAttentionAllYou2023}, allowing the model to pay attention to all users at once, learning cross-user contention patterns that drive how the RAN allocates resources (\cref{fig:neural_scheduler_architecture}).
   
\subparahead{MCS model:} This model selects the MCS---unlike PRB allocation, the MCS mainly depends on an individual user's specific CQI, which is largely independent of other users in the cell. Because it does not need to model UE competition, its design is more direct, using transformer encoder layers to track individual CQI over time, followed by fully-connected layers to predict the correct MCS target, yielding a similar structure as the \trafficmodel{} in \cref{fig:traffic_reconstructor}.

\parahead{Synthesizing the transmission budget.} Finally, the output of these two 
models---PRB allocation and MCS index---are combined to compute the transmission budgets (TBS) for each user over the coming slots.
By continuously predicting and enforcing the budget, the \schedulermodel{}, integrated into the emulation system (\S\ref{s:impl}), faithfully recreates the highly dynamic, workload-dependent scheduler of a live 5G network.

\begin{figure}
    \centering
    \includegraphics[width=\linewidth]{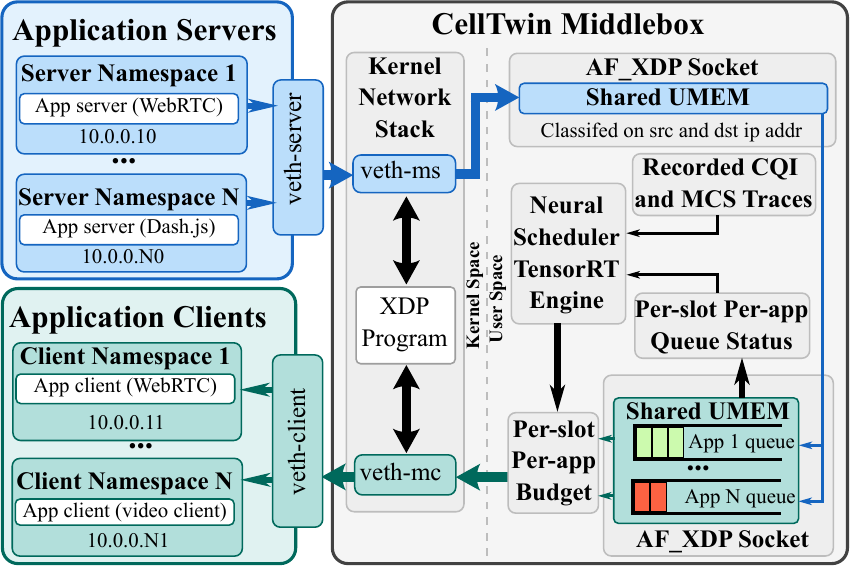}
    \caption{\sysname{}'s emulation system implementation.}
    \label{fig:emu_sys}
    \vskip -0.15in
\end{figure}

\section{Implementation}\label{s:impl}
\parahead{Emulation system.}
We implement the \sysname{} emulation system as a high-performance middlebox utilizing 
the Linux \texttt{AF\_XDP} library (\cref{fig:emu_sys}). 
We employ two cross-linked sockets to manage bi-directional traffic. 
During the emulation, the \schedulermodel{} inputs recorded CQI traces 
from live users or traces, and the buffer 
size in the middlebox to make a dynamic scheduling decision.
At every slot boundary, the system replenishes the transmission budget based on the \schedulermodel{}'s output. 
Furthermore, the system enforces a TDD UL/DL slot pattern, zeroing the downlink budget during uplink-heavy slots to mimic the frame structure of production 5G Standalone (SA).
The system mimics the HARQ retransmission mechanism of the 5G network:
when the packet is lost at a given loss rate, the network retries after ten slots, crucial to replicate 
RAN delay jitter.

\sysname{}'s inference and deployment code is written in around 2K lines of C++ code.
We deploy the \schedulermodel{} using the NVIDIA TensorRT framework for low inference latency. 
The models are quantized to FP16 precision and executed on a dedicated GPU thread to avoid blocking the main packet-processing loop. 
We utilize pre-allocated GPU buffers and asynchronous memory copies to minimize data transfer overhead, ensuring that scheduling decisions for the next 10 slots are always available to the main loop.

\parahead{Training data collection.}
To expose more information from the live 5G network, we engineer a randomized data collection framework.
We use four laptops tethered to Samsung S22 Ultra smartphones as our controlled users, and one GCP VM as the application server.
We organize the data collection with 2-minute sessions, and the data has a high resolution -- 2000 slots (0.5 ms) per second, making it 240K time-series data points per session per user.
For each data collecting session, we spawn random number of users from 1 to 4, and for each user, we select a random application and setup listed in \cref{tab:data_app}. 
The data collection spans across three months from January to April 2026 and we collected 237 data sessions in total. 
We summarize the user number (controlled and background) and MCS distributions in \cref{fig:cellular-status} (Appendix \ref{appx:cell_status}).

\parahead{ML model training.} 
\sysname{} uses PyTorch distributed training framework to speed up the training process. 
We train all four models from scratch on A100 GPUs through slurm job and each training task takes one week to converge.
The models take input with granularity of 5G slots, yielding 2000 samples per second from collected data.
The data are divided with the ratio of 8:1:1 for training, validation and test sets. 
\sysname{}'s checkpoint contains 3.2 and 1.1 MB models in \schedulermodel{} for PRB allocation and MCS selection.

\begin{figure}
    \centering
    \includegraphics[width=\linewidth]{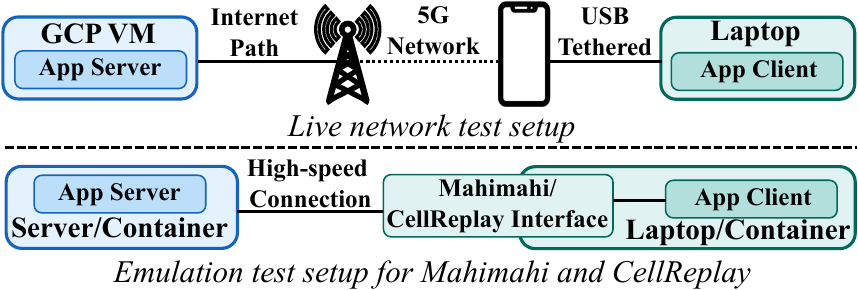}
    \caption{Test setup for live 5G, Mahimahi and CellReplay.}
    \label{fig:eval_setup}
    \vskip -0.15in
\end{figure}


\section{Evaluation}

Our goal is to evaluate \sysname{}'s emulation accuracy with CellReplay\cite{sentosaCellReplayAccurateRecordandreplay2025} and Mahimahi\cite{netravali_mahimahi_2015} in approximating application performance compared to its live 5G network counterpart.
%

\subsection{Methodology}
\parahead{Evaluation setup.}
We design the evaluation setup for live 5G network, CellReplay and Mahimahi, as shown in \cref{fig:eval_setup}. 
\sysname{} uses the same setup, but it's deployed in a separate entity as a middlebox program, as shown in \cref{fig:emu_sys}.
The live network setup is used to test application performance in the live 5G network.
The live network is a T-Mobile standalone 5G network, operating in band 41 with 2.5 GHz center frequency and 100 MHz bandwidth. 
It offers up to 400 Mbps total capacity (with multiple users), and single user can achieve 270 Mbps through a saturator. 
During the tests, we tethered a laptop to a phone connected to 5G network, and we set up the application server at GCP, at a distance of 350 miles, with a wired RTT of 32 ms, to imitate the real application performance.
For the emulation setup, we deploy docker containers for application server and client, then we apply CellReplay and Mahimahi interface on the client to shape the traffic.
For \sysname{}, we deploy another container to serve as the middlebox and set up multiple network namespaces in both ends to emulate multiple concurrent traffic flows, as shown in \cref{fig:emu_sys}.

\begin{figure}
    \centering
    \subfigure[Congestion control throughput.]{
    \includegraphics[width=0.49\linewidth]{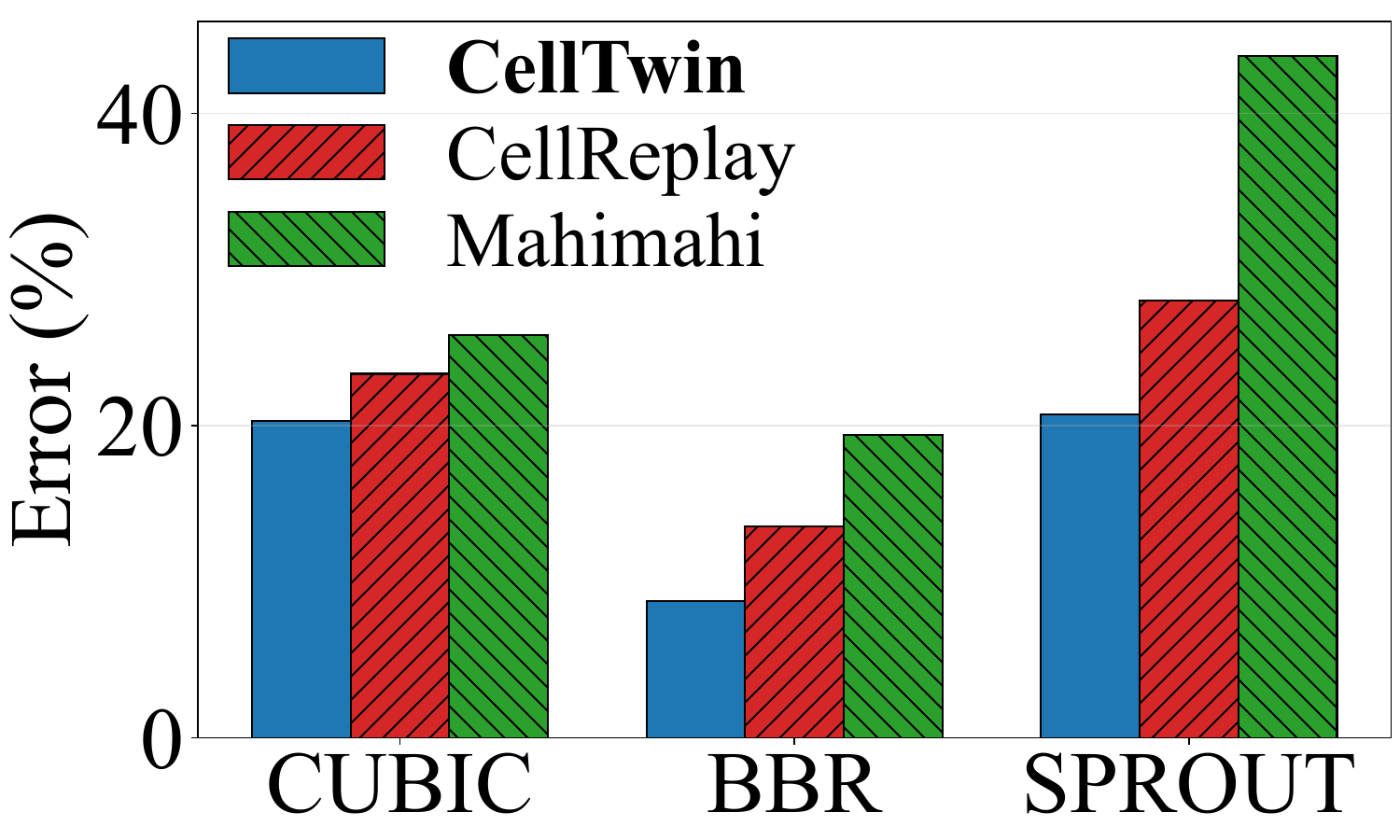}\label{fig:single_tcp_tput}}
    \hfill
    \subfigure[Congestion control delay.]
    {\includegraphics[width=0.49\linewidth]{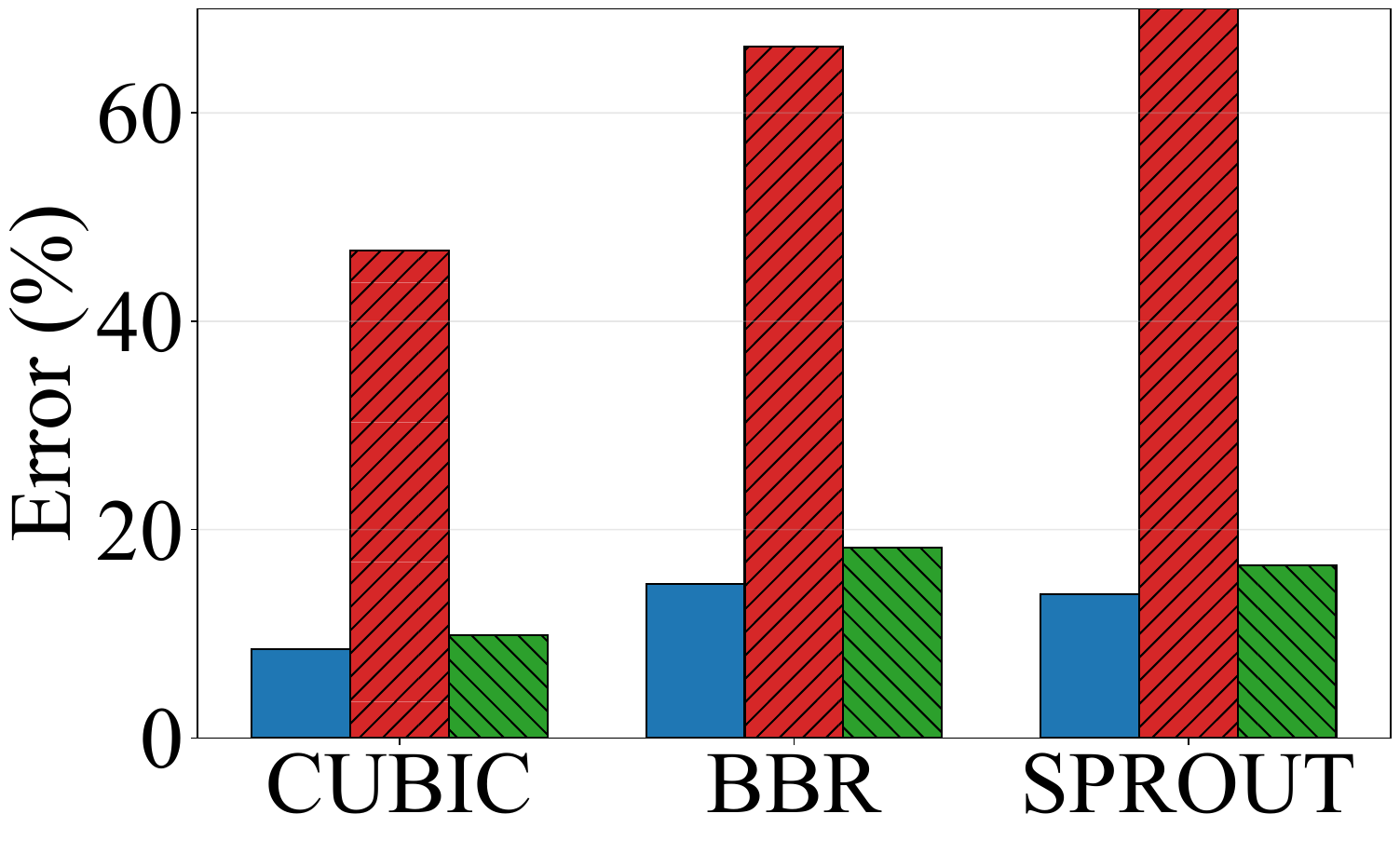}\label{fig:single_tcp_owd}}
    \caption{Congestion control algorithms' \textit{emulation distribution errors} in the single-user scenario.}
    \label{fig:single_tcp} 
    \vskip -0.25in
\end{figure}

For all the evaluations, we interleave the data collection for ground truth and channel data.
We first run the application in the live 5G network, during which we collect the user's CQI and MCS traces with CellNinja\cite{liuSeeingFogEmpowering2025}, then right after, we record the CellReplay and Mahimahi traces, so that the RAN status (user contention and channel status) is similar.
We repeat this interleaving process for five times for data collection in a short time before we start the evaluation, so that the ground truth and the trace data share similar network conditions.
With these traces, we run the same application in the emulation settings and compare the results.
We compare the emulation error with CellReplay and Mahimahi under single-user scenarios, because they only support the single-user scenario.
We also demonstrate \sysname{}'s ability in a multi-user scenario, which is the first of its kind.

\parahead{Emulation error metrics.}
We quantify the emulation errors using the normalized distributions difference between network environments: one from the live 5G network and the other from the emulation. 
We use Earth Mover's Distance (EMD) \cite{rubner_emd_1998}, defined as: $\mathrm{EMD}(L, T) = \int_{-\infty}^{\infty} |L(x) - T(x)|dx$, where $L$ and $T$ are the CDF of two distributions, in our case, the observed application metrics from live 5G network and emulation.
A lower EMD indicates a closer distance between the two distributions, hence better fidelity. 
We calculate \textit{emulation distribution error} in percentage by dividing the EMD 
with the mean metrics value from the live network tests.

\parahead{Tested applications:}
We test the same application categories listed in \cref{tab:data_app}. 
Within each category, we include unseen cases and behaviors to test the model's generalizability.

\begin{figure}
    \centering
    \subfigure[HTTP/1.1]{
    \includegraphics[width=0.49\linewidth]{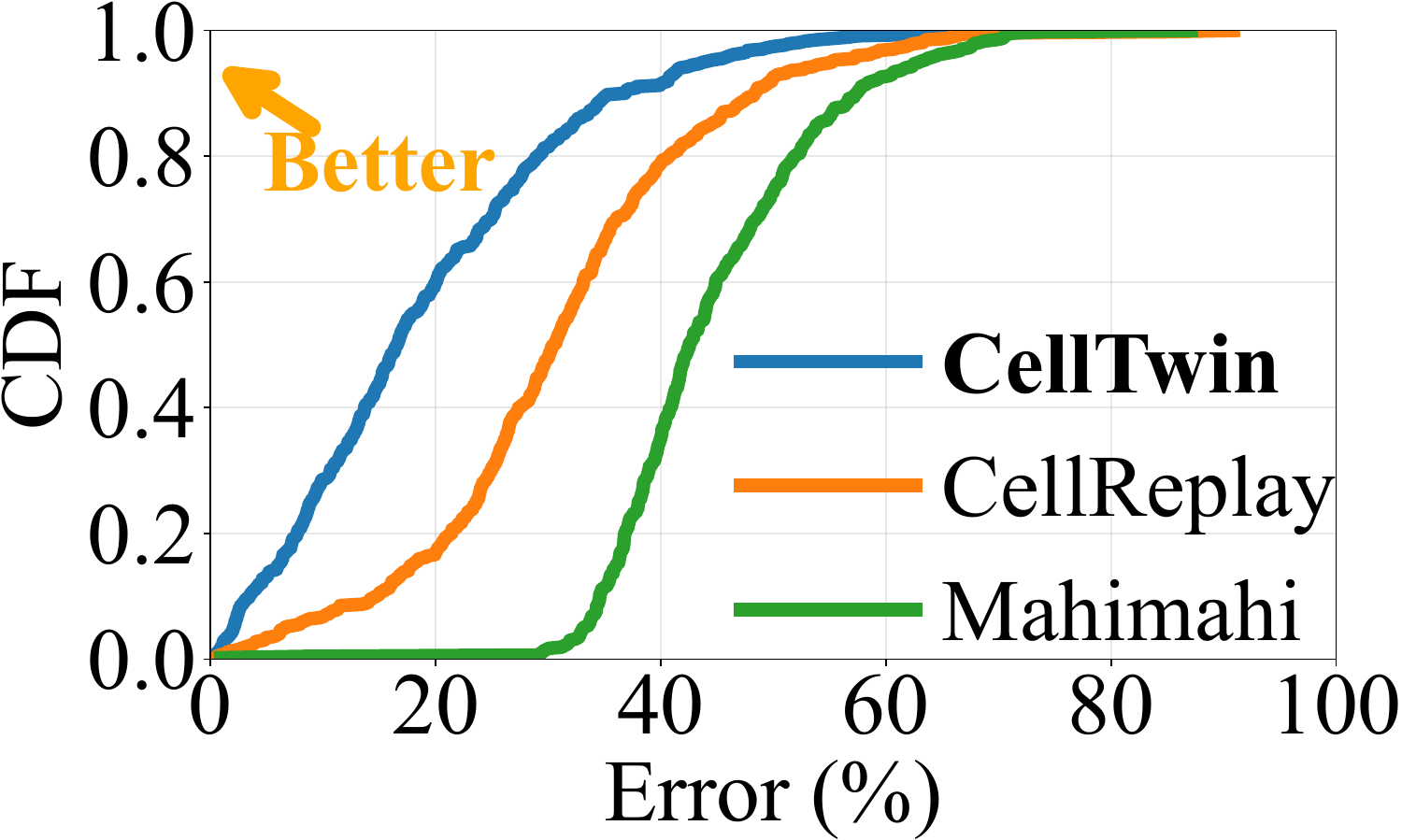}\label{fig:http1-1}}
    \hfill
    \subfigure[HTTP/2]
    {\includegraphics[width=0.49\linewidth]{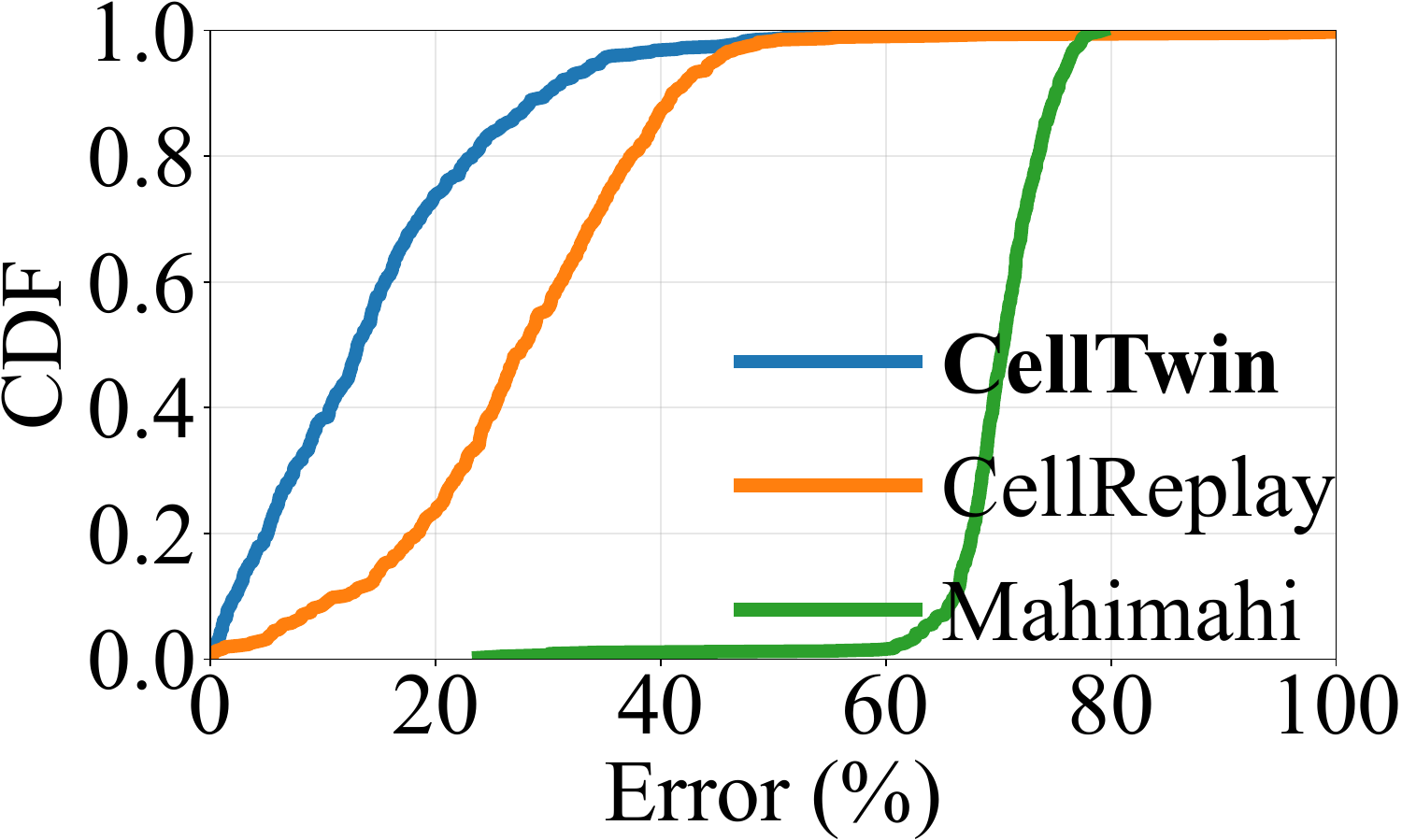}\label{fig:http2}}
    \caption{Webpage load time results of \sysname{}, CellReplay and Mahimahi, compared with the live network.}
    \label{fig:single_webpage} 
    \vskip -0.15in
\end{figure}




\subsection{Single-user Emulation}\label{s:eval:single_user}

\parahead{Congestion control algorithm's performance.}
Here we run different congestion control algorithms in the live and emulation environments. 
We include the widely used CUBIC \cite{ha_cubic_2008} and BBR \cite{cardwell_bbr_2016} schemes in the Linux kernel and evaluate research community developed schemes, including Sprout \cite{winstein_stochastic_2013}, whose behaviors are not seen by our model during training.
We run each of these schemes in the live network for 60 seconds, then collect the traces for the emulators, and run them in the emulators.
We repeat this procedure for five times and summarize the emulation error results.
As shown in \cref{fig:single_tcp}, \sysname{} reduces the \textit{emulation distribution error} across all three congestion control schemes, in throughput and delay metrics.
Surprisingly, CellReplay has the worst performance in delay emulation results in this setup.
The root cause could be that their light-PDO design adds extra delays into the queuing delay during the emulation.
In the throughput emulation metrics, the emulation schemes have a closer gap and \sysname{} has the lowest emulation errors among them.
We exhibit a time-series Copa performance in Appendix \ref{appx:copa}.

\begin{figure}
    \centering
    \subfigure[VoD bit rate selection (kbps).]{
    \includegraphics[width=0.49\linewidth]{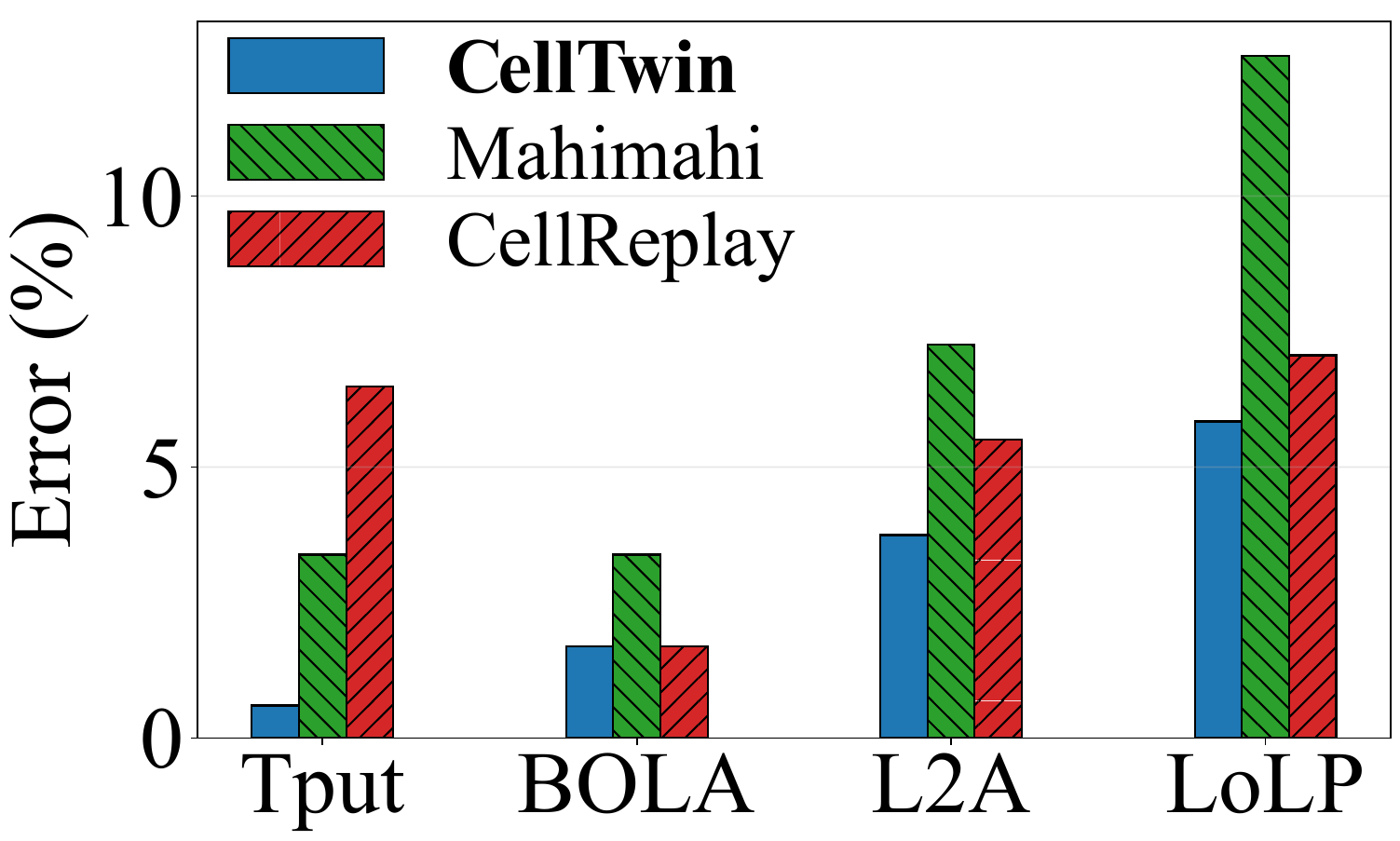}\label{fig:single_vod_br}}
    \hfill
    \subfigure[VoD buffer level (seconds).]
    {\includegraphics[width=0.49\linewidth]{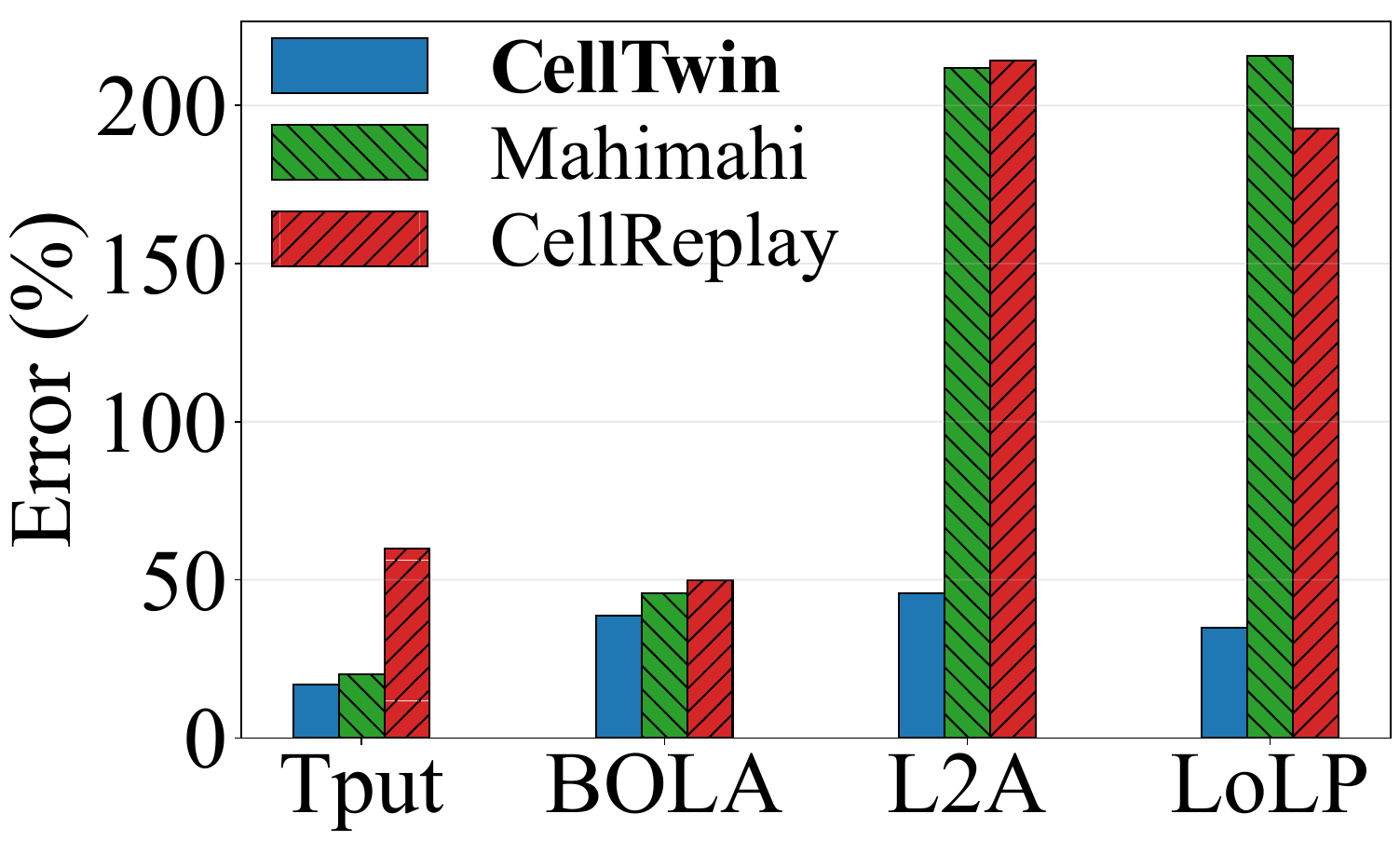}\label{fig:single_vod_buffer}}
    \caption{Single user VoD application \textit{emulation distribution errors} in bit rate selection and playback buffer size.}
    \label{fig:single_vod} 
    \vskip -0.15in
\end{figure}

\parahead{Web page load time.}
Here we evaluate \sysname{} in web page browsing application. 
We serve 94 Wikipedia pages in the server with the same script in both live 5G network and the emulations.
%
We load these webpages with actual headless chrome browser, controlled by a python script, and measured page-load time (PLT), based on the
\texttt{Page.loadEventFired} event, directly corresponding to the standard JavaScript \texttt{window.onload} event, as the metrics.
%
We test both HTTP/1.1 and HTTP/2 and repeat the experiments for 5 times, then we calculate the PLT absolute error in percentage for every page.
\cref{fig:single_webpage} shows the results, where the average web page loading time is 929~ms in the live network.
\sysname{} outperforms CellReplay and Mahimahi in both HTTP/1.1 and HTTP/2 scenarios. 
\sysname{} has an average error reduction of 12\% and 25\% in HTTP/1.1 scenario and average error reduction of 13\% and 55\% in HTTP/2 scenario, relative to CellReplay and Mahimahi, respectively.
Both CellReplay and Mahimahi underestimate web page load time, because the sizes are relatively small and they both use a saturated network trace (heavy PDO) for the replay, yielding an overestimated throughput. 
%

\parahead{Video-on-demand application performance.}
To evaluate the performance of the video-on-demand application, we select the ABR algorithms implemented in Dash.js \cite{noauthor_dash_nodate} that don't appear in our data collection. 
The data is collected with the default Dynamic ABR algorithm, which will dynamically adapt between throughput-based (Tput for short) rule and the buffer-based rule BOLA \cite{spiteri_bola_2020}.
In the evaluation, we select the ABR algorithms of individual Throughput-based and BOLA, as well as L2A \cite{karagkioules_l2a_2022}---a learning based scheme designed for mobile network, and LoL+ \cite{lolp_lim_2020}, which is designed for low latency live streaming.
The video client eventually reaches the maximum bit rate, given the 270 Mbps single-user throughput, so similarly as in CellReplay \cite{sentosaCellReplayAccurateRecordandreplay2025}, we capture the initial phase (first 30 seconds) of the video client before it fully ramps onto the 15Mbps bit rate.
We repeat the video application for five times under each environment, and \cref{fig:single_vod} shows the results of this evaluation in bit rate selection and the playback buffer level in seconds.
\sysname{}'s \textit{emulation distribution errors} for bit rate selection remains under 5$\%$ for throughput, BOLA and L2A schemes, while it has a higher error in emulating LoL+'s bit rate selection (5.9\%).
For playback buffer level, \sysname{} has higher emualation errors, ranging from 15\% to 48\%, but it still performs the best among all three.
\begin{figure}
    \centering
    \subfigure[WebRTC encoder target bit rate.]{
    \includegraphics[width=0.49\linewidth]{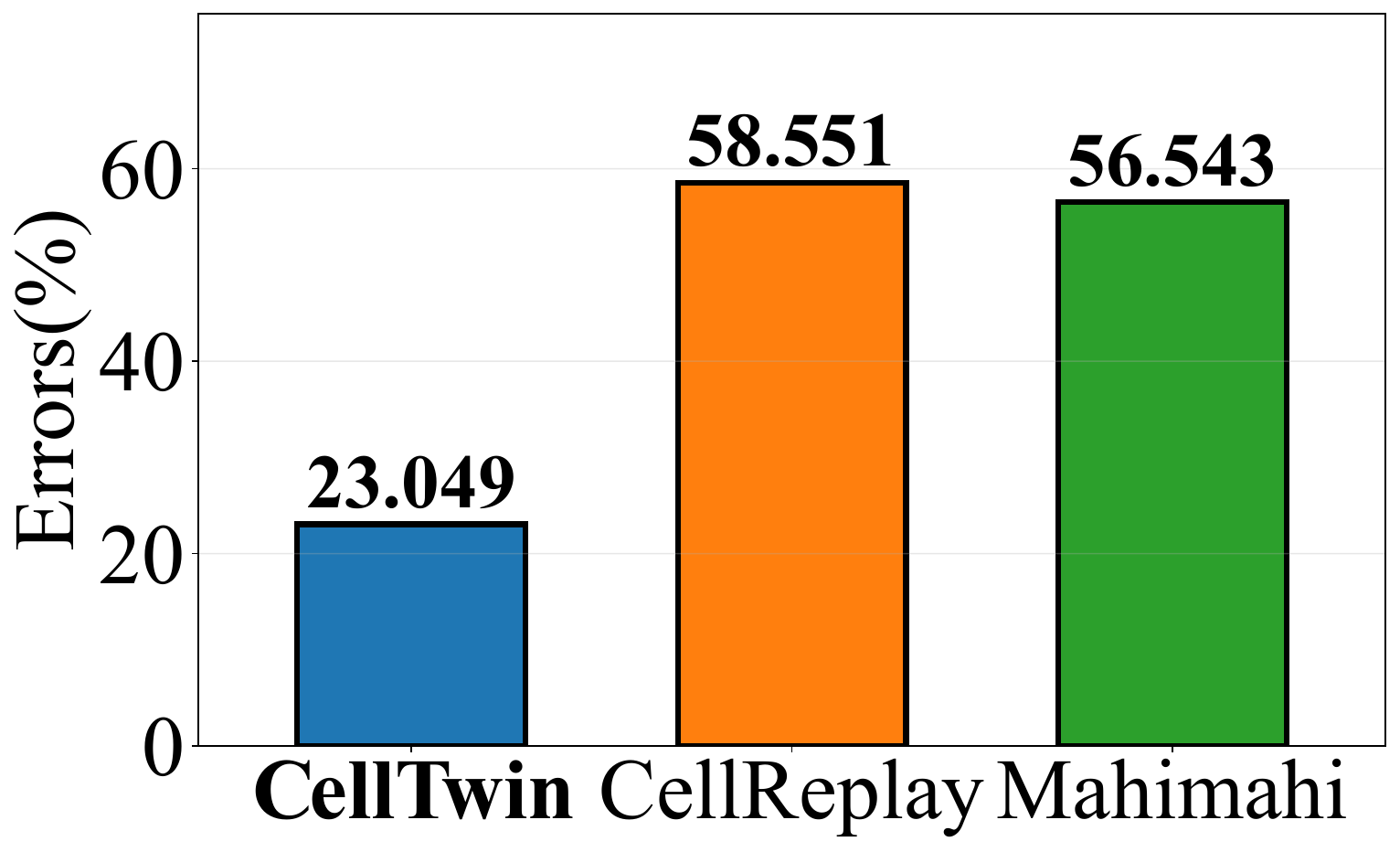}\label{fig:single_webrtc_bitrate}}
    \hfill
    \subfigure[RTT perceived by WebRTC.]
    {\includegraphics[width=0.49\linewidth]{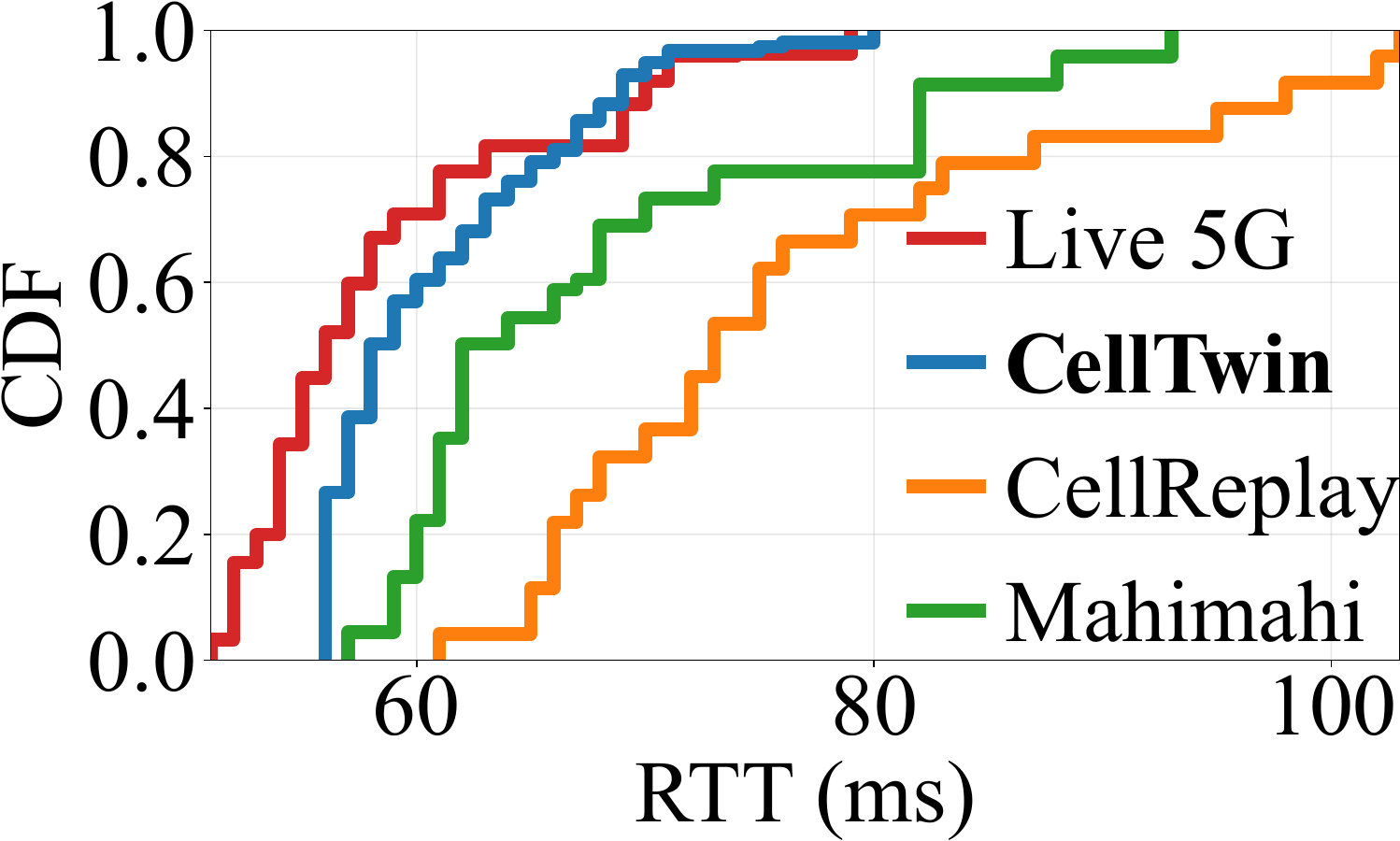}\label{fig:single_webrtc_rtt}}
    \caption{Single user WebRTC emulation results in target bit rate and application-perceived RTT.}
    \label{fig:single_webrtc} 
\end{figure}

\begin{figure}
    \centering
    \includegraphics[width=\linewidth]{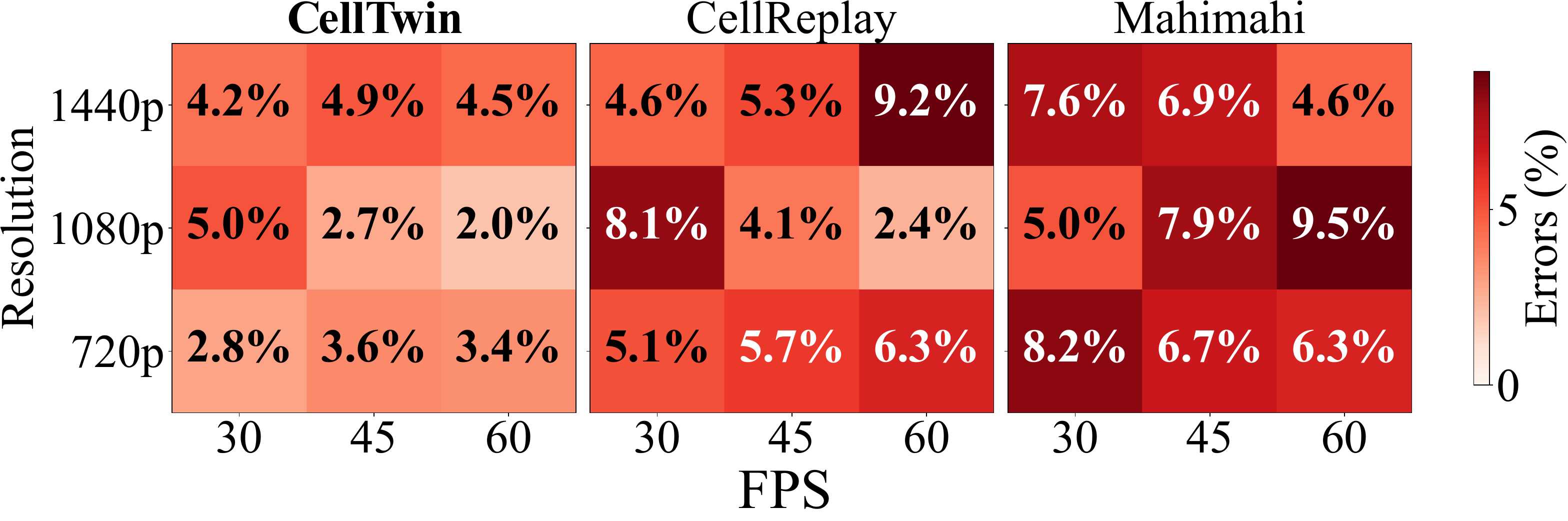}
    \caption{Single user cloud gaming \textit{emulation distribution errors} in client-perceived delays.}
    \label{fig:single_cloudgaming}
    \vskip -0.15in
\end{figure}

\parahead{WebRTC evaluation.}
For WebRTC, we set up the peer-to-peer video conferencing standalone application directly from its C++ library.
We use the default GCC \cite{blum_webrtc_2021} for the bit rate selection, which monitors the delay variations in the network to determine the "overuse" and "underuse" internal states and adjusts the target bit rate for the encoder.
We evaluate the \textit{emulation distribution errors} of the encoder target bit rate, which changes more frequently than VoD application, due to its real-time nature.
We extract the application internal metrics following \cite{yiAutomatedCrossLayerRoot2025}, and report the application's encoder target bit rate and perceived RTT in \cref{fig:single_webrtc}.
\sysname{} has a target bit rate emulation error of 23\%, while both CellReplay and Mahimahi's errors exceed 55\%.
As for the app perceived RTT, \sysname{} is closest to the live 5G network counterpart.

\parahead{Cloud gaming.}
The open source cloud gaming application software Sunshine server\cite{lizardbyte_sunshine_2023} and Moonlight client\cite{gutman_moonlight_2023} don't have an ABR algorithm, relying on the user's configuration instead.
For the evaluation, we sweep across different resolutions (720p to 1440p) and frame rate configurations (30 to 60 frames/sec)
and compare the perceived delay distribution reported by the client.
\cref{fig:single_cloudgaming} shows the results, where \sysname{} achieves the lowest \textit{emulation distribution errors}.

\subsection{Multi-user Emulation}

\begin{figure}
    \begin{minipage}[b]{0.475\linewidth}
        \centering
        \includegraphics[width=\linewidth]{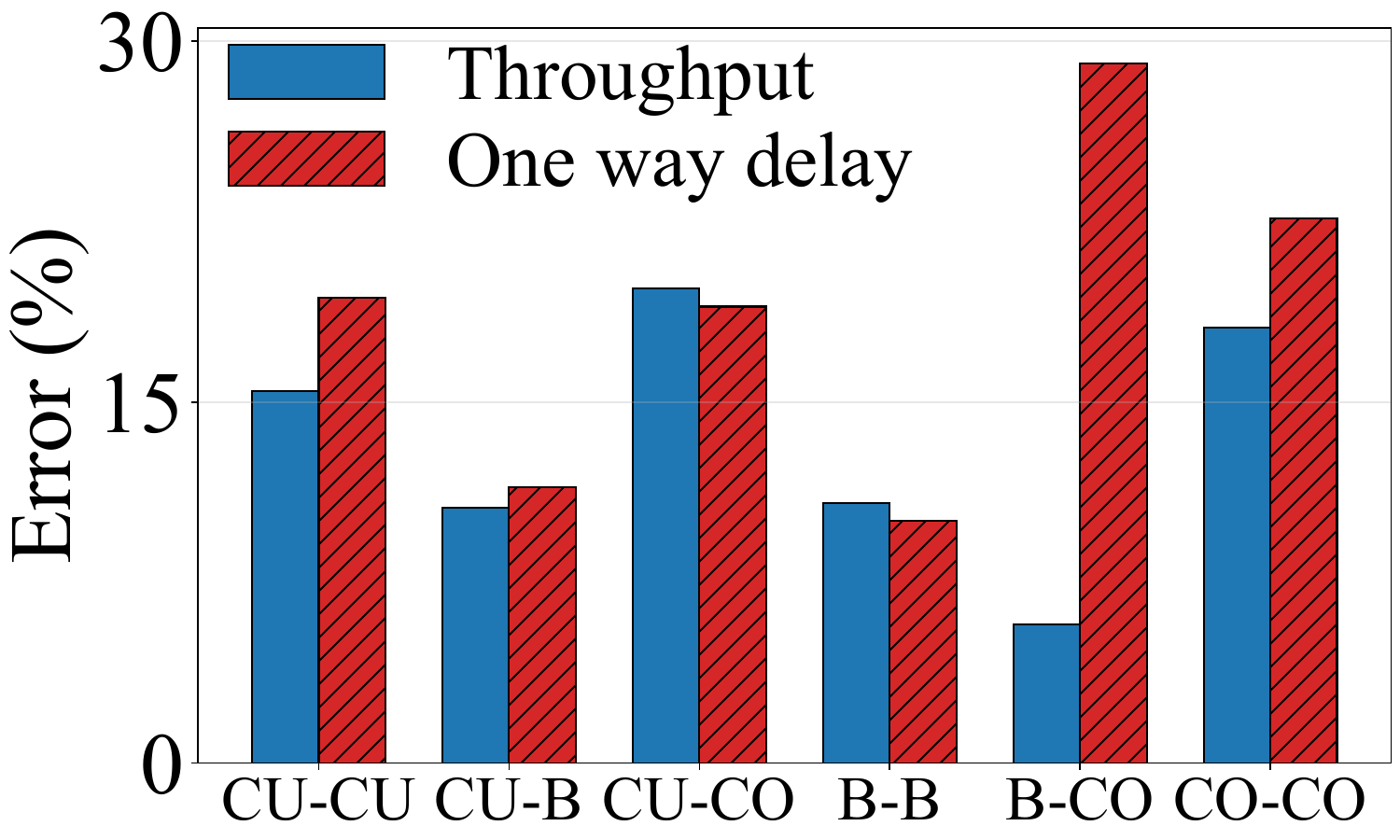}
        \caption{\textit{Emulation distribution errors} for two congestion control combinations.}
    \label{fig:2user_cc}
    \end{minipage}
    \hfill
    \begin{minipage}[b]{0.475\linewidth}          
       \centering
        \includegraphics[width=\linewidth]{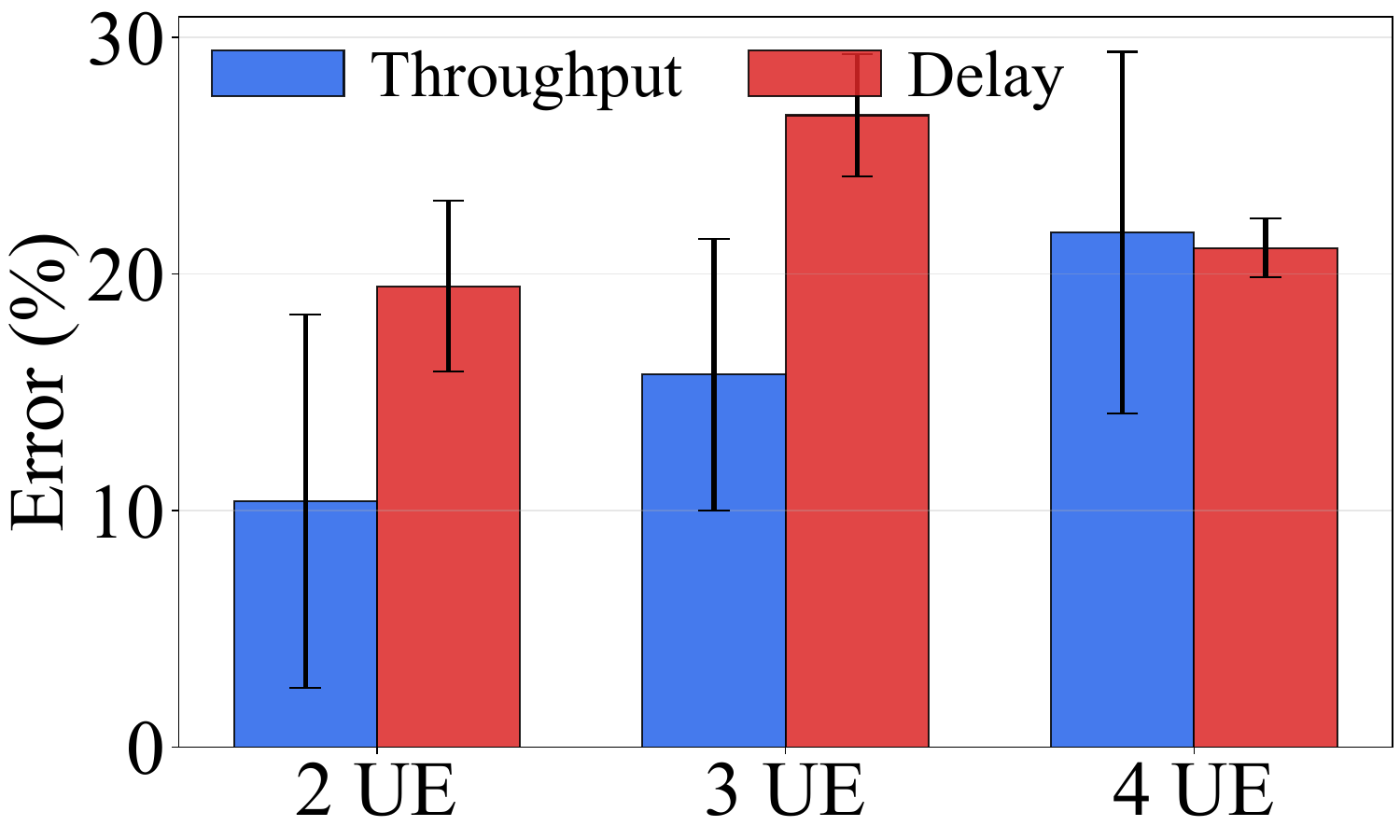}
        \caption{\textit{Merged EMD errors} for concurrent multi-UE mixed applications.}
        \label{fig:multi_mix}   
    \end{minipage}
    \vskip -0.25in
\end{figure}

\parahead{Concurrent TCP flows.}
To evaluate \sysname{}'s emulation performance in multi-user scenario, we initiate two concurrent flows into the live 5G network and \sysname{} emulation. 
Here we try to emulate the interactions between congestion control algorithms, including \underline{CU}BIC, \underline{B}BR and \underline{Co}pa, denoted as CU, B, and CO in the figure respectively.
For both live 5G network and \sysname{} emulation, we run \texttt{iperf3} for TCP schemes and collect the packet traces with \texttt{tcpdump} for throughput and RTT calculation.
We run Copa with its customized UDP implementation, and we set up tunnels on both ends to calculate the ingress throughput and delay metrics.
We run each combination for 2 minutes and repeat the experiment for five times.
We calculate the \textit{emulation distribution errors} of the throughput and RTT/delay for each flow in the combinations.
\cref{fig:2user_cc} shows the results, across most combinations, \sysname{} achieves emulation distribution errors around 20\%. 

\begin{figure*}
    \centering
    \includegraphics[width=\linewidth]{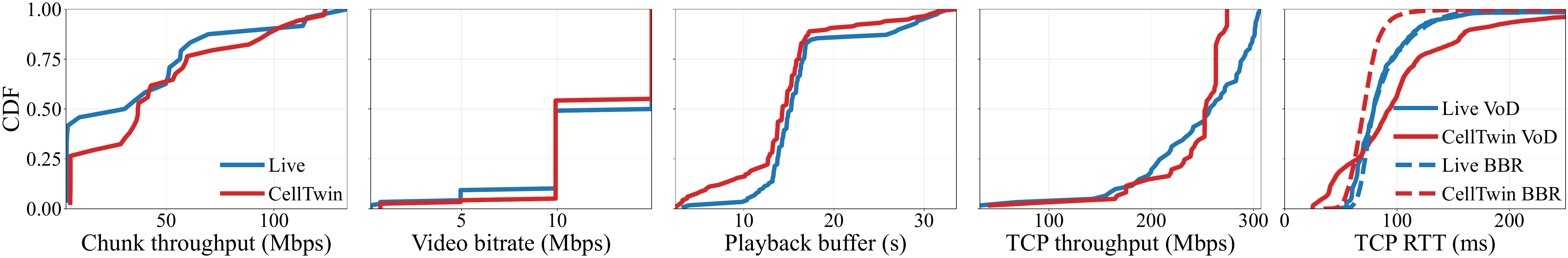}
    \caption{Evaluation result between live 5G and \sysname{}, where a VoD flow is sharing the RAN resources with a TCP BBR flow.}
    \label{fig:l2a_vs_bbr}
    \vskip -0.15in
\end{figure*}

\parahead{Multi-UE Mixed Applications.}
Here we evaluate \schedulermodel{}'s performance with multiple concurrent UEs and mixed applications. 
We start with 2 UEs with TCP and VoD flows, then we add one UE at a time with a different application, for 3 UEs, the additional application is webpage loading, and we add the cloud gaming application for 4 UEs setting. 
Then we calculate the throughput and latency distribution for all the different applications, and compare them against the live 5G ground truth to calculate the emulation distribution errors.
For each setting, we repeat the experiment for 5 times and merge the distribution error results.
\cref{fig:multi_mix} shows the results, \schedulermodel{} achieves 19.5\%, 26.7\% and 21.1\% average delay emulation errors, and 10.4\%, 15.8\% and 21.7\% average throughput emulation errors.

As a continuation of the previous discussion on the RAN sharing dynamics (\S\ref{s:motivation:failure}), we show the detailed emulation performance of VoD ABR algorithm (L2A\cite{karagkioules_l2a_2022}) with a BBR background TCP flow.
%
%
%
\cref{fig:l2a_vs_bbr} shows the CDF comparison between the live 5G network and \sysname{} emulation in five application metrics, including the VoD per-chunk throughput, video bit rate selection, playback buffer length, the competing TCP flow's throughput, and the RTTs of the VoD and BBR.
In all five metrics, \sysname{}'s emulation distributions are very close to the live network's counterpart, as opposed to \cref{fig:contention_emulation}.


\begin{figure}
    \begin{minipage}[b]{0.475\linewidth}
        \centering
        \includegraphics[width=\linewidth]{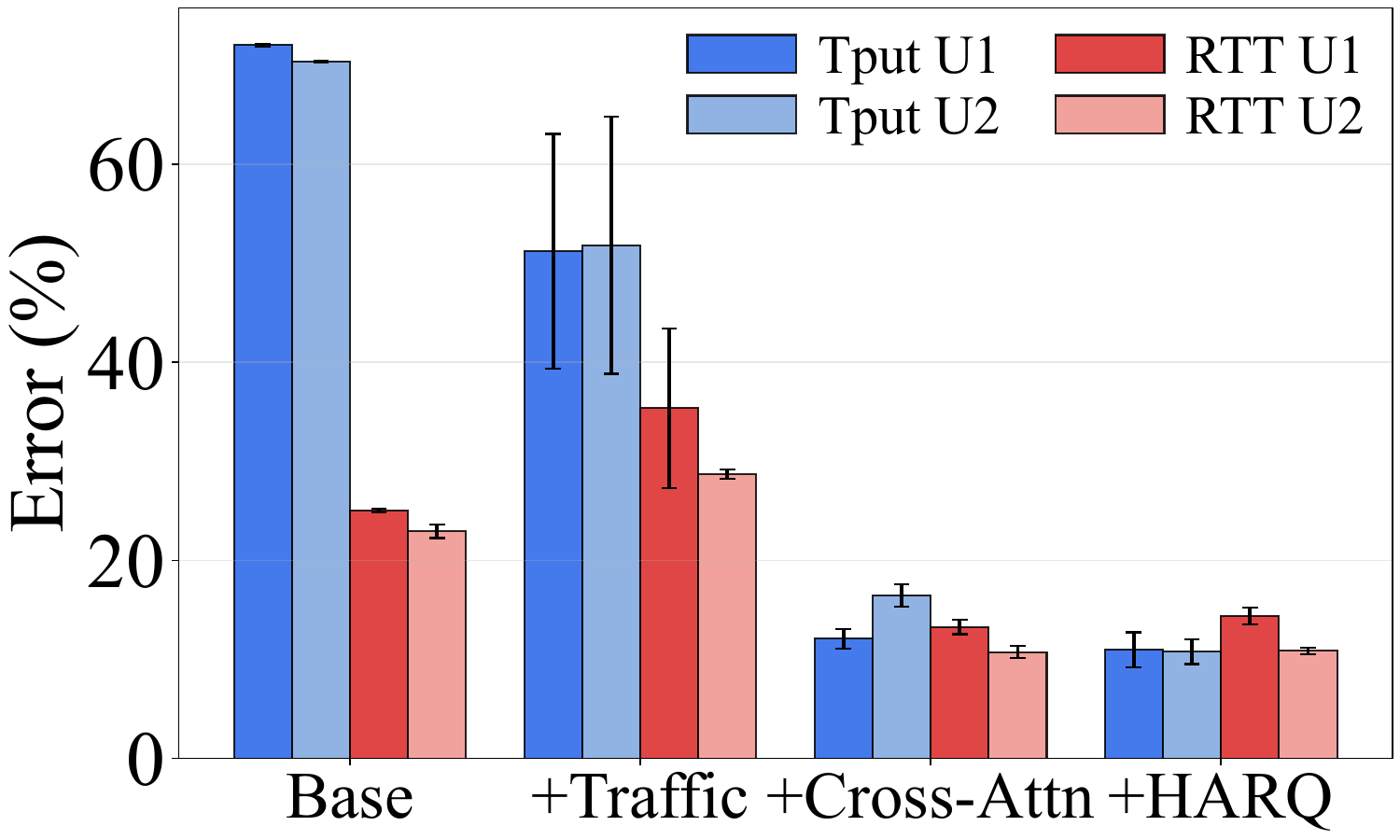}
        \caption{Ablation study for the design components.}
    \label{fig:module_abl}
    \end{minipage}
    \hfill
    \begin{minipage}[b]{0.475\linewidth}          
       \centering
        \includegraphics[width=\linewidth]{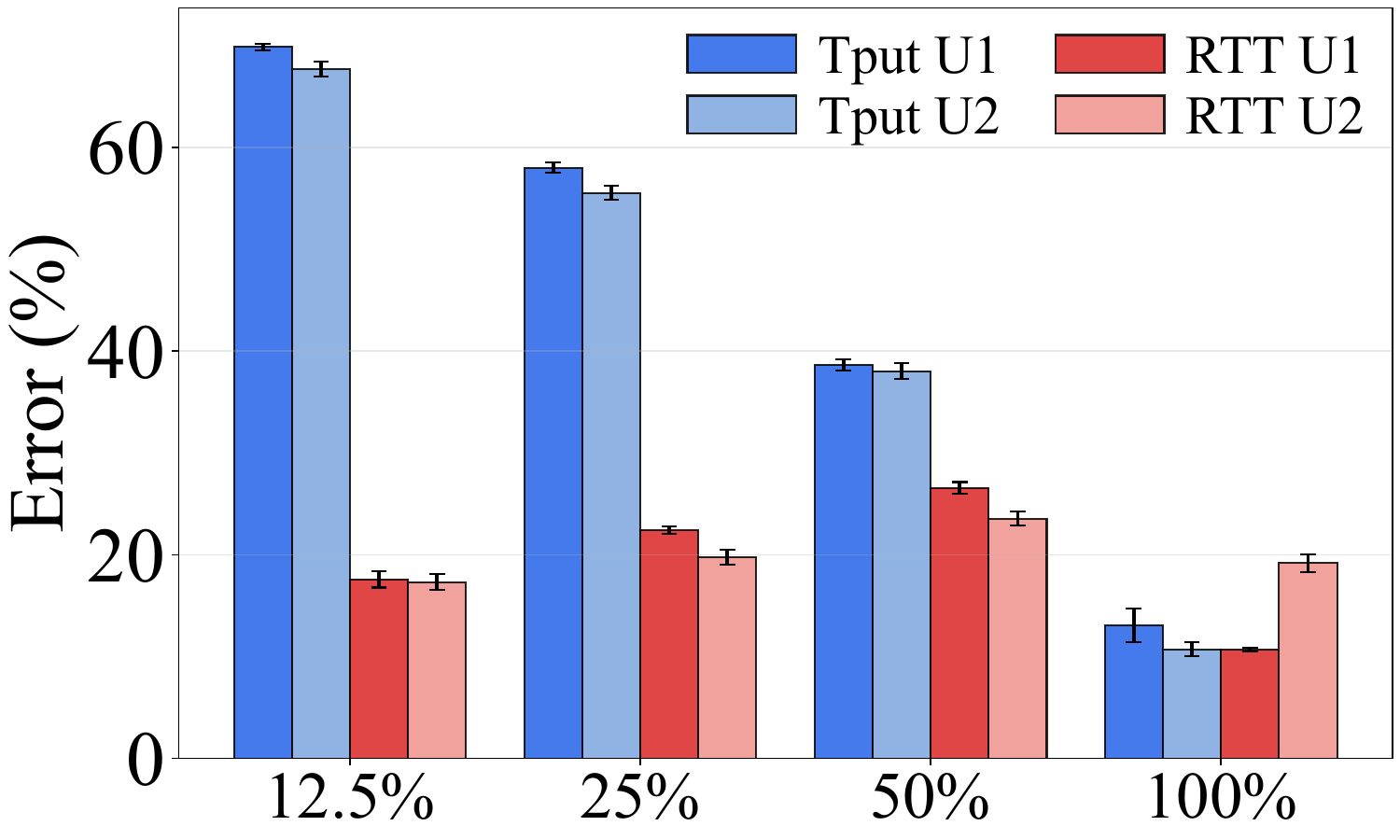}
        \caption{Ablation study for dataset size in portion.}
        \label{fig:data_abl}   
    \end{minipage}
    \vskip -0.25in
\end{figure}

\subsection{Microbenchmarks}

\parahead{Traffic Reconstructor.} Here we evaluate the performance of \trafficmodel{} on a test set it never sees during training, which learns the reverse mapping from the scheduling results to the incoming packet pattern and CQI metrics.
%
%
%
For the incoming packet count, the reconstructed per-slot value matches ground truth 
within ±1 packet for 54\% of slots, within ±3 for 80\%, and within ±4 for 90\%; the estimator is nearly unbiased, under-counting by only 0.15 packets/slot. 
For CQI, the model recovers the exact index in 76\% of active slots and falls within one CQI level in 97\% (within two levels in 99.6\%), with a negligible bias of +0.07 CQI, showing minimal offset.
%

\parahead{Module Ablation Study.}
Here we conduct an ablation study to evaluate the effectiveness of each design component, including the \trafficmodel{}, cross-user attention and the HARQ emulation in the middlebox system by re-training the \schedulermodel{} without these components with same setting and training epochs. 
We use the two-UE TCP BBR flows as the case study, \cref{fig:module_abl} shows the results of this evaluation.
\trafficmodel{} can help to lower the error from base model, and the cross-user attention module further captures the scheduler's behavior with concurrent flows.
HARQ emulation helps to mitigate the tail distribution errors.

\parahead{Data Ablation Study.}
Here we evaluate how the dataset size helps to improve the performance, we train the model with different portion (uniformly sampled 12.5\%, 25\%, 50\% and 100\%) of the dataset with the same setting and epoch numbers, and evaluate the model's performance with two concurrent BBR flows.
\cref{fig:data_abl} shows the result -- more data exposes more behavior variability and yields better performance.

\section{Related Work}
\parahead{Network emulators.}
NetEm\cite{HemmingerNetworkEW} is a Linux kernel enhancement within the \texttt{tc} subsystem, which can be used to emulate network links with delay variations.
Google Chrome provides network throttle profile with fixed bandwidth and latency for cellular network emulation \cite{chrome_devtools_network_throttling}.
Mahimahi\cite{netravali_mahimahi_2015} is a framework for recording and replaying HTTP traffic, and it includes a network emulator derived from CellSim \cite{winstein_stochastic_2013} for cellular link emulation.
Pantheon\cite{yan_pantheon_2018} is an algorithm evaluation framework, in which two endpoints deployed UDP tunnels for bit rate calculation and time synchronization, and it provides handy wrappers for evaluating network algorithms for the community.
NemFi\cite{mishra_NemFi_2021} develops a record-and-replay emulator solely for Wi-Fi network. 
CellReplay\cite{sentosaCellReplayAccurateRecordandreplay2025} introduces a dual-workload trace structure, trying to capture the work-load dependent behavior of the cellular network.
However, all the schemes discussed only capture a fixed snapshot of the network under a specific scenario, while \sysname{} learns the core scheduling intelligence of the cellular network. 

\parahead{RAN Digital Twin.}
Another line of work tries to build an authentic radio environment emulator for the RAN.
Geo2SigMap\cite{li2024geo2sigmap} tries to estimate propagation path loss with deep learning models for precise radio map generation.
ChannelGAN\cite{xiao_channelgan_2022} aims channel state information emulation with generative adversarial networks \cite{goodfellow2014generative}.
RadioTwin\cite{zhenlin_radiotwin_2025} builds a physically interpretable digital twin of the real-world radio environment.
These methods are strictly physical-layer emulators; consequently, a massive gap remains for researchers aiming to evaluate higher-layer network algorithms and transport protocols, which is the target of \sysname{}.


\parahead{Learning-based RAN control.}
A parallel line of work uses ML to design MAC-layer decisions rather than to reproduce them. 
Q-learning \cite{9024384} and decoupled DRL on a USRP prototype \cite{you2026simulationrealitypracticaldeep} replace the CQI-to-MCS look-up table and outer-loop link adaptation with learned policies, while DRL agents have been trained to jointly select users and contiguous RBs in 5G scheduling \cite{sun2020deepreinforcementlearningbasedschedulingcontiguousresource} and to allocate slice-aware RAN resources on a live O-RAN testbed \cite{rezazadeh2023multiagentdeepreinforcementlearning}. 
CellTwin instead treats the deployed commercial RAN scheduler as the learning target, approximating its
behavior for network emulation.

\section{Limitations and Future Work}

\parahead{Mobility.}
Our evaluation uses real indoor static and moving user's channels (Appendix \ref{appx:mobility}). 
%
%
In handover, the buffered data are sent from one RAN to another, one can use two \sysname{} and a handover emulator \cite{hassan_vivisecting_2022} for the emulation.

\parahead{Uplink emulation.} In this work, we focus on downlink data transmission.
RANs may use different uplink scheduling algorithms than the downlink, so retraining on uplink is necessary for a good emulation performance.
Fortunately, UEs send their
buffer status to the base station, which 
can be extracted by CellNinja, obviating the need for
the buffer estimation (\S\ref{s:design:data_collection}).
Then the training and model design can be reused.

\parahead{Emulating different vendors.}
We focus on T-Mobile but our design
flow works universally across vendors and operators. 
It's necessary to retrain the model for a different vendor/operator or when the RAN changes its scheduling algorithm.
For the training, we trained our four models on four A100 GPUs individually for one week for the reported performance.
For data collection, one needs phones tethered to laptops and connected to the target operator, and one SDR for RAN data collection, GPS-driven NTP clocks for precise timing and a cloud server for the traffic.
The data, model, and automatic data collection 
framework are released for the research community to build 
their own \sysname{} for different vendors.

\section{Conclusion}


A commercial RAN's capacity is not a property of the link but the output of a proprietary scheduler reacting to every user in the cell.
Record-and-replay freezes that reaction; full-stack simulators reproduce every layer except the logic that matters. \sysname{} instead learns the scheduler itself, recovering the buffer and channel state the base station never exposes. 
Running as a lightweight middlebox, it roughly halves emulation error across web, video, conferencing, and gaming, and is the first emulator to place multiple flows in genuine contention. 

\section*{Acknowledgements}
This material is based upon work supported by the National Science Foundation under grants AST-2232457, CNS-2223556, and CNS-2433915.

\newpage
\let\oldbibliography\thebibliography
\renewcommand{\thebibliography}[1]{%
  \oldbibliography{#1}%
  \setlength{\parskip}{0pt}%
  \setlength{\itemsep}{0pt}%
}
\bibliographystyle{concise2}
\bibliography{references}

@misc{pultarova_5g_2026,
	title = {{5G} {Networks} {Offer} {Secure} {GPS} {Alternatives}},
    note = {{IEEE} {Spectrum}},
	howpublished = {\href{https://spectrum.ieee.org/5g-gnss-gps-alternatives}{spectrum.ieee.org}},
    author = {Pultarova, Tereza},
	language = {en},
	date = {2026-04-21},
}

@misc{ConfigureNTPCompute,
	title = {{Google Compute Engine:} {Configure} {NTP} on a compute instance},
    key = {Google},
	howpublished = {\href{https://docs.cloud.google.com/compute/docs/instances/time-synchronization/configure-ntp}{docs.cloud.google.com}},
	language = {en},
	urldate = {2026-04-09},
	note = {Google Cloud Documentation},
}

@misc{PublicNTP,
	title = {Google Public {NTP}},
	howpublished = {\href{https://developers.google.com/time}{developers.google.com}},
    key = {Google},
	language = {en},
	urldate = {2026-04-09},
	note = {Google for Developers},
}

@misc{vaswaniAttentionAllYou2023,
	title = {Attention {Is} {All} {You} {Need}},
	url = {http://arxiv.org/abs/1706.03762},
	doi = {10.48550/arXiv.1706.03762},
	urldate = {2026-04-09},
	publisher = {arXiv},
	author = {Vaswani, Ashish and Shazeer, Noam and Parmar, Niki and Uszkoreit, Jakob and Jones, Llion and Gomez, Aidan N. and Kaiser, Lukasz and Polosukhin, Illia},
	month = aug,
	year = {2023},
	note = {arXiv:1706.03762 [cs]},
}

@inproceedings{choLearningPhraseRepresentations2014,
	address = {Doha, Qatar},
	title = {Learning {Phrase} {Representations} using {RNN} {Encoder}–{Decoder} for {Statistical} {Machine} {Translation}},
	url = {https://aclanthology.org/D14-1179/},
	doi = {10.3115/v1/D14-1179},
	urldate = {2026-04-09},
	booktitle = {Proceedings of the 2014 {Conference} on {Empirical} {Methods} in {Natural} {Language} {Processing} ({EMNLP})},
	publisher = {Association for Computational Linguistics},
	author = {Cho, Kyunghyun and van Merriënboer, Bart and Gulcehre, Caglar and Bahdanau, Dzmitry and Bougares, Fethi and Schwenk, Holger and Bengio, Yoshua},
	editor = {Moschitti, Alessandro and Pang, Bo and Daelemans, Walter},
	month = oct,
	year = {2014},
	pages = {1724--1734},
}

@article{kushnerConvergenceProportionalfairSharing2004,
	title = {Convergence of proportional-fair sharing algorithms under general conditions},
	volume = {3},
	issn = {1558-2248},
	url = {https://ieeexplore.ieee.org/document/1310314},
	doi = {10.1109/TWC.2004.830826},
	number = {4},
	urldate = {2026-04-04},
	journal = {IEEE Transactions on Wireless Communications},
	author = {Kushner, H.J. and Whiting, P.A.},
	month = jul,
	year = {2004},
	pages = {1250--1259},
}

@book{arpaci-dusseauOperatingSystemsThree2018,
	edition = {1.00},
	title = {Operating {Systems}: {Three} {Easy} {Pieces}},
	publisher = {Arpaci-Dusseau Books},
	author = {Arpaci-Dusseau, Remzi H. and Arpaci-Dusseau, Andrea C.},
	month = aug,
	year = {2018},
}

@inproceedings{10.1145/3097766.3097772,
author = {Mahfoudi, Mohamed Naoufal and Turletti, Thierry and Parmentelat, Thierry and Dabbous, Walid},
title = {Lessons Learned while Trying to Reproduce the OpenRF Experiment},
year = {2017},
isbn = {9781450350600},
publisher = {Association for Computing Machinery},
address = {New York, NY, USA},
url = {https://doi.org/10.1145/3097766.3097772},
doi = {10.1145/3097766.3097772},
booktitle = {Proceedings of the Reproducibility Workshop},
pages = {21–23},
numpages = {3},
location = {Los Angeles, CA, USA},
series = {Reproducibility '17}
}

@inproceedings{chang:can-you-see-me,
  author = {Chang, Hyunseok and Varvello, Matteo and Hao, Fang and Mukherjee, Sarit},
  title = {Can You See Me Now? A Measurement Study of Zoom, Webex, and Meet},
  year = {2021},
  isbn = {9781450391290},
  publisher = {ACM},
  address = {New York, NY, USA},
  url = {https://doi.org/10.1145/3487552.3487847},
  booktitle = {ACM Internet Measurement Conference},
  pages = {216--228}
}

@inproceedings{varvello:vca-performance-in-wild,
  author = {Varvello, Matteo and Chang, Hyunseok and Zaki, Yasir},
  title = {Performance characterization of videoconferencing in the wild},
  year = {2022},
  isbn = {9781450392594},
  publisher = {Association for Computing Machinery},
  address = {New York, NY, USA},
  url = {https://doi.org/10.1145/3517745.3561442},
  doi = {10.1145/3517745.3561442},
  booktitle = {Proceedings of the 22nd ACM Internet Measurement Conference},
  pages = {261--273},
  numpages = {13},
  location = {Nice, France},
  series = {IMC '22}
}

@article{10.1145/972374.972386,
author = {Judd, Glenn and Steenkiste, Peter},
title = {Repeatable and realistic wireless experimentation through physical emulation},
year = {2004},
issue_date = {January 2004},
publisher = {Association for Computing Machinery},
address = {New York, NY, USA},
volume = {34},
number = {1},
issn = {0146-4833},
url = {https://doi.org/10.1145/972374.972386},
doi = {10.1145/972374.972386},
journal = {SIGCOMM Comput. Commun. Rev.},
month = jan,
pages = {63–68},
numpages = {6}
}

@inproceedings {246302,
author = {Prateesh Goyal and Anup Agarwal and Ravi Netravali and Mohammad Alizadeh and Hari Balakrishnan},
title = {{ABC}: A Simple Explicit Congestion Controller for Wireless Networks },
booktitle = {17th USENIX Symposium on Networked Systems Design and Implementation (NSDI 20)},
year = {2020},
isbn = {978-1-939133-13-7},
address = {Santa Clara, CA},
pages = {353--372},
url = {https://www.usenix.org/conference/nsdi20/presentation/goyal},
publisher = {USENIX Association},
month = feb
}

@inproceedings {211251,
author = {Sadjad Fouladi and John Emmons and Emre Orbay and Catherine Wu and Riad S. Wahby and Keith Winstein},
title = {Salsify: {Low-Latency} Network Video through Tighter Integration between a Video Codec and a Transport Protocol},
booktitle = {15th USENIX Symposium on Networked Systems Design and Implementation (NSDI 18)},
year = {2018},
isbn = {978-1-939133-01-4},
address = {Renton, WA},
pages = {267--282},
url = {https://www.usenix.org/conference/nsdi18/presentation/fouladi},
publisher = {USENIX Association},
month = apr
}

@ARTICLE{10472093,
  author={Li, Xishuo and Zhang, Shan and Huang, Yuejiao and Ma, Xiao and Wang, Zhiyuan and Luo, Hongbin},
  journal={IEEE Transactions on Mobile Computing}, 
  title={Towards Timely Video Analytics Services at the Network Edge}, 
  year={2024},
  volume={23},
  number={11},
  pages={10443-10459},
  doi={10.1109/TMC.2024.3376769}}

@inproceedings{10.1145/3769102.3770618,
author = {Ghasemi, Mahshid and Fu, Yongjie and Ouyang, Xinyu and Wang, Peiran and Turkcan, Mehmet Kerem and Tavori, Jhonatan and Kleisarchaki, Sofia and Calmant, Thomas and G\"{u}rgen, Levent and Kostic, Zoran and Di, Xuan Sharon and Zussman, Gil and Ghaderi, Javad},
title = {Real-Time Video Analytics for Urban Safety: Deployment over Edge and End Devices},
year = {2025},
isbn = {9798400722387},
publisher = {Association for Computing Machinery},
address = {New York, NY, USA},
url = {https://doi.org/10.1145/3769102.3770618},
doi = {10.1145/3769102.3770618},
booktitle = {Proceedings of the Tenth ACM/IEEE Symposium on Edge Computing},
articleno = {14},
numpages = {17},
location = {the Hilton Arlington National Landing, Arlington, VA, USA},
series = {SEC '25}
}

@inproceedings{10.1145/3307334.3328589,
author = {Ananthanarayanan, Ganesh and Bahl, Victor and Cox, Landon and Crown, Alex and Nogbahi, Shadi and Shu, Yuanchao},
title = {Video Analytics - Killer App for Edge Computing},
year = {2019},
isbn = {9781450366618},
publisher = {Association for Computing Machinery},
address = {New York, NY, USA},
url = {https://doi.org/10.1145/3307334.3328589},
doi = {10.1145/3307334.3328589},
booktitle = {Proceedings of the 17th Annual International Conference on Mobile Systems, Applications, and Services},
pages = {695–696},
numpages = {2},
location = {Seoul, Republic of Korea},
series = {MobiSys '19}
}

@article{yu:mustang,
  author = {Yu, Encheng and Zhou, Jianer and Li, Zhenyu and Tyson, Gareth and Li, Weichao and Zhang, Xinyi and Xu, Zhiwei and Xie, Gaogang},
  title = {Mustang: Improving QoE for Real-Time Video in Cellular Networks by Masking Jitter},
  year = {2024},
  issue_date = {September 2024},
  publisher = {Association for Computing Machinery},
  address = {New York, NY, USA},
  volume = {20},
  number = {9},
  issn = {1551-6857},
  url = {https://doi.org/10.1145/3672399},
  doi = {10.1145/3672399},
  journal = {ACM Trans. Multimedia Comput. Commun. Appl.},
  month = sep,
  articleno = {285},
  numpages = {23}
}

@misc{OaiOpenairinterface5GGitLab2026,
	title = {{OAI} / {openairinterface5G} {5G Wireless Implementation}},
	howpublished = {\href{https://gitlab.eurecom.fr/oai/openairinterface5g}{eurecom.fr}},
	language = {en},
	urldate = {2026-04-04},
	note = {GitLab},
    key = {OAI},
	month = apr,
	year = {2026},
}

@misc{liuSeeingFogEmpowering2025,
	title = {Seeing {Through} the {Fog}: {Empowering} {Mobile} {Devices} to {Expose} and {Mitigate} {RAN} {Buffer} {Effects} on {Delay}-{Sensitive} {Protocols}},
	shorttitle = {Seeing {Through} the {Fog}},
	url = {https://arxiv.org/abs/2507.00337v2},
	language = {en},
	urldate = {2026-04-03},
	journal = {arXiv.org},
	author = {Liu, Yuxin and Zhang, Tianyang and Jamieson, Kyle and Xie, Yaxiong},
	month = jul,
	year = {2025},
}

@article{zhangArsenalUnderstandingLearningBased2021,
	title = {Arsenal: {Understanding} {Learning}-{Based} {Wireless} {Video} {Transport} via {In}-{Depth} {Evaluation}},
	volume = {70},
	issn = {1939-9359},
	shorttitle = {Arsenal},
	url = {https://ieeexplore.ieee.org/abstract/document/9516990},
	doi = {10.1109/TVT.2021.3105479},
	number = {10},
	urldate = {2026-04-03},
	journal = {IEEE Transactions on Vehicular Technology},
	author = {Zhang, Huanhuan and Zhou, Anfu and Ma, Ruoxuan and Lu, Jiamin and Ma, Huadong},
	month = oct,
	year = {2021},
	pages = {10832--10844},
}

@inproceedings{yuMagpieImprovingEfficiency2024,
	address = {New York, NY, USA},
	series = {{IMC} '24},
	title = {Magpie: {Improving} the {Efficiency} of {A}/{B} {Tests} for {Large} {Scale} {Video}-on-{Demand} {Systems}},
	isbn = {979-8-4007-0592-2},
	shorttitle = {Magpie},
	url = {https://dl.acm.org/doi/10.1145/3646547.3689019},
	doi = {10.1145/3646547.3689019},
	urldate = {2026-04-02},
	booktitle = {Proceedings of the 2024 {ACM} on {Internet} {Measurement} {Conference}},
	publisher = {Association for Computing Machinery},
	author = {Yu, Hebin and Wang, Haiping and Tian, Chenfei and Sathyanarayana, Sandesh Dhawaskar and Shi, Shu and Xue, Zhichen and Yu, Shuaixin and Li, Haozhe and Peng, Yajie and Pang, Xiaofei and Zhang, Ruixiao},
	month = nov,
	year = {2024},
	pages = {588--594},
}

@misc{nsnamNs3,
	title = {ns-3: a discrete-event network simulator for internet systems},
	howpublished = {\href{https://www.nsnam.org}{nsnam.org}},
	language = {en},
	urldate = {2026-04-03},
	journal = {ns-3},
	author = {nsnam},
}

@inproceedings{sentosaCellReplayAccurateRecordandreplay2025,
	title = {{CellReplay}: {Towards} accurate record-and-replay for cellular networks},
	isbn = {978-1-939133-46-5},
	shorttitle = {{CellReplay}},
    booktitle = {USENIX NSDI},
	url = {https://www.usenix.org/conference/nsdi25/presentation/sentosa},
	language = {en},
	urldate = {2026-04-03},
	author = {Sentosa, William and Chandrasekaran, Balakrishnan and Godfrey, P. Brighten and Hassanieh, Haitham},
	year = {2025},
	pages = {1169--1186},
}

@inproceedings{yiAutomatedCrossLayerRoot2025,
	address = {New York, NY, USA},
	series = {{IMC} '25},
	title = {Automated, {Cross}-{Layer} {Root} {Cause} {Analysis} of {5G} {Video}-{Conferencing} {Quality} {Degradation}},
	isbn = {979-8-4007-1860-1},
	url = {https://dl.acm.org/doi/10.1145/3730567.3764434},
	doi = {10.1145/3730567.3764434},
	urldate = {2026-04-02},
	booktitle = {Proceedings of the 2025 {ACM} {Internet} {Measurement} {Conference}},
	publisher = {Association for Computing Machinery},
	author = {Yi, Fan and Wan, Haoran and Jamieson, Kyle and Michel, Oliver},
	month = nov,
	year = {2025},
	pages = {835--850},
}

@misc{liLargeModelEnabled2025,
	title = {Large {Model} {Enabled} {Embodied} {Intelligence} for {6G} {Integrated} {Perception}, {Communication}, and {Computation} {Network}},
	url = {http://arxiv.org/abs/2512.15109},
	doi = {10.48550/arXiv.2512.15109},
	urldate = {2026-04-03},
	publisher = {arXiv},
	author = {Li, Zhuoran and Gao, Zhen and Liu, Xinhua and Wang, Zheng and Zhou, Xiaotian and Liu, Lei and Wu, Yongpeng and Feng, Wei and Huang, Yongming},
	month = dec,
	year = {2025},
	note = {arXiv:2512.15109 [eess]
version: 1},
}

@article{SynergeticEmpowermentWireless,
      title={{Synergetic Empowerment: Wireless Communications Meets Embodied Intelligence}}, 
      author={Hongtao Liang and Yihe Diao and YuHang Wu and Fuhui Zhou and Qihui Wu},
      year={2025},
      number={2509.10481},
      journal={arXiv},
      primaryClass={cs.NI},
      howpublished={https://arxiv.org/abs/2509.10481}, 
}

@inproceedings{liDissectingStreamliningInteractive2025,
	title = {Dissecting and {Streamlining} the {Interactive} {Loop} of {Mobile} {Cloud} {Gaming}},
	isbn = {978-1-939133-46-5},
	url = {https://www.usenix.org/conference/nsdi25/presentation/li-yang},
	language = {en},
	urldate = {2026-04-03},
	author = {Li, Yang and Qiu, Jiaxing and Wang, Hongyi and Li, Zhenhua and Qian, Feng and Yang, Jing and Lin, Hao and Liu, Yunhao and Xiao, Bo and Qin, Xiaokang and Xu, Tianyin},
	year = {2025},
	pages = {595--611},
}

@inproceedings{yi_athena_2024,
	address = {New York, NY, USA},
	series = {{HotNets} '24},
	title = {Athena: {Seeing} and {Mitigating} {Wireless} {Impact} on {Video} {Conferencing} and {Beyond}},
	isbn = {9798400712722},
	shorttitle = {Athena},
	url = {https://dl.acm.org/doi/10.1145/3696348.3696889},
	doi = {10.1145/3696348.3696889},
	urldate = {2025-05-30},
	booktitle = {Proceedings of the 23rd {ACM} {Workshop} on {Hot} {Topics} in {Networks}},
	publisher = {Association for Computing Machinery},
	author = {Yi, Fan and Wan, Haoran and Jamieson, Kyle and Rexford, Jennifer and Xie, Yaxiong and Michel, Oliver},
	month = nov,
	year = {2024},
	pages = {103--110},
}

@inproceedings{wan_nr-scope_2024,
	address = {New York, NY, USA},
	series = {{CoNEXT} '24},
	title = {{NR}-{Scope}: {A} {Practical} {5G} {Standalone} {Telemetry} {Tool}},
	isbn = {9798400711084},
	shorttitle = {{NR}-{Scope}},
	url = {https://dl.acm.org/doi/10.1145/3680121.3697808},
	doi = {10.1145/3680121.3697808},
	urldate = {2025-05-30},
	booktitle = {Proceedings of the 20th {International} {Conference} on emerging {Networking} {EXperiments} and {Technologies}},
	publisher = {Association for Computing Machinery},
	author = {Wan, Haoran and Cao, Xuyang and Marder, Alexander and Jamieson, Kyle},
	month = dec,
	year = {2024},
	pages = {73--80},
}

@inproceedings{ye_dissecting_2024,
	address = {New York, NY, USA},
	series = {{ACM} {SIGCOMM} '24},
	title = {Dissecting {Carrier} {Aggregation} in {5G} {Networks}: {Measurement}, {QoE} {Implications} and {Prediction}},
	isbn = {9798400706141},
	shorttitle = {Dissecting {Carrier} {Aggregation} in {5G} {Networks}},
	url = {https://dl.acm.org/doi/10.1145/3651890.3672250},
	doi = {10.1145/3651890.3672250},
	urldate = {2025-01-31},
	booktitle = {Proceedings of the {ACM} {SIGCOMM} 2024 {Conference}},
	publisher = {Association for Computing Machinery},
	author = {Ye, Wei and Hu, Xinyue and Sleder, Steven and Zhang, Anlan and Dayalan, Udhaya Kumar and Hassan, Ahmad and Fezeu, Rostand A. K. and Jajoo, Akshay and Lee, Myungjin and Ramadan, Eman and Qian, Feng and Zhang, Zhi-Li},
	month = aug,
	year = {2024},
	pages = {340--357},
}

@misc{wan_evolving_2024,
	title = {Evolving {Mobile} {Cloud} {Gaming} with {5G} {Standalone} {Network} {Telemetry}},
	url = {http://arxiv.org/abs/2402.04454},
	doi = {10.48550/arXiv.2402.04454},
	urldate = {2024-08-06},
	publisher = {arXiv},
	author = {Wan, Haoran and Jamieson, Kyle},
	month = feb,
	year = {2024},
	note = {arXiv:2402.04454 [cs]},
}

@article{ha_cubic_2008,
	title = {{CUBIC}: a new {TCP}-friendly high-speed {TCP} variant},
	volume = {42},
	issn = {0163-5980},
	shorttitle = {{CUBIC}},
	url = {https://dl.acm.org/doi/10.1145/1400097.1400105},
	doi = {10.1145/1400097.1400105},
	number = {5},
	urldate = {2024-05-03},
	journal = {ACM SIGOPS Operating Systems Review},
	author = {Ha, Sangtae and Rhee, Injong and Xu, Lisong},
	month = jul,
	year = {2008},
	pages = {64--74},
}

@article{blum_webrtc_2021,
	title = {{WebRTC}: real-time communication for the open web platform},
	volume = {64},
	issn = {0001-0782, 1557-7317},
	shorttitle = {{WebRTC}},
	url = {https://dl.acm.org/doi/10.1145/3453182},
	doi = {10.1145/3453182},
	language = {en},
	number = {8},
	urldate = {2024-05-03},
	journal = {Communications of the ACM},
	author = {Blum, Niklas and Lachapelle, Serge and Alvestrand, Harald},
	month = aug,
	year = {2021},
	pages = {50--54},
}

@article{xie_ng-scope_2022,
	title = {{NG}-{Scope}: {Fine}-{Grained} {Telemetry} for {NextG} {Cellular} {Networks}},
	volume = {6},
	issn = {2476-1249},
	shorttitle = {{NG}-{Scope}},
	url = {https://dl.acm.org/doi/10.1145/3508032},
	doi = {10.1145/3508032},
	language = {en},
	number = {1},
	urldate = {2024-05-03},
	journal = {Proceedings of the ACM on Measurement and Analysis of Computing Systems},
	author = {Xie, Yaxiong and Jamieson, Kyle},
	month = feb,
	year = {2022},
	pages = {1--26},
}

@inproceedings{hassan_vivisecting_2022,
	address = {New York, NY, USA},
	series = {{SIGCOMM} '22},
	title = {Vivisecting mobility management in {5G} cellular networks},
	isbn = {9781450394208},
	url = {https://dl.acm.org/doi/10.1145/3544216.3544217},
	doi = {10.1145/3544216.3544217},
	urldate = {2024-04-23},
	booktitle = {Proceedings of the {ACM} {SIGCOMM} 2022 {Conference}},
	publisher = {Association for Computing Machinery},
	author = {Hassan, Ahmad and Narayanan, Arvind and Zhang, Anlan and Ye, Wei and Zhu, Ruiyang and Jin, Shuowei and Carpenter, Jason and Mao, Z. Morley and Qian, Feng and Zhang, Zhi-Li},
	month = aug,
	year = {2022},
	pages = {86--100},
}

@inproceedings{arun_copa_2018,
	address = {Montreal QC Canada},
	title = {Copa: {Practical} {Delay}-{Based} {Congestion} {Control} for the {Internet}},
	isbn = {978-1-4503-5585-8},
	shorttitle = {Copa},
	url = {https://dl.acm.org/doi/10.1145/3232755.3232783},
	doi = {10.1145/3232755.3232783},
	language = {en},
	urldate = {2024-04-23},
	booktitle = {Proceedings of the {Applied} {Networking} {Research} {Workshop}},
	publisher = {ACM},
	author = {Arun, Venkat and Balakrishnan, Hari},
	month = jul,
	year = {2018},
	pages = {19--19},
}

@article{cardwell_bbr_2016,
	title = {{BBR}: {Congestion}-{Based} {Congestion} {Control}: {Measuring} bottleneck bandwidth and round-trip propagation time},
	volume = {14},
	issn = {1542-7730},
	shorttitle = {{BBR}},
	url = {https://dl.acm.org/doi/10.1145/3012426.3022184},
	doi = {10.1145/3012426.3022184},
	number = {5},
	urldate = {2024-04-22},
	journal = {Queue},
	author = {Cardwell, Neal and Cheng, Yuchung and Gunn, C. Stephen and Yeganeh, Soheil Hassas and Jacobson, Van},
	month = oct,
	year = {2016},
	pages = {20--53},
}

@article{spiteri_bola_2020,
	title = {{BOLA}: {Near}-{Optimal} {Bitrate} {Adaptation} for {Online} {Videos}},
	volume = {28},
	issn = {1063-6692},
	shorttitle = {{BOLA}},
	url = {https://dl.acm.org/doi/10.1109/TNET.2020.2996964},
	doi = {10.1109/TNET.2020.2996964},
	number = {4},
	urldate = {2024-01-29},
	journal = {IEEE/ACM Transactions on Networking},
	author = {Spiteri, Kevin and Urgaonkar, Rahul and Sitaraman, Ramesh K.},
	month = aug,
	year = {2020},
	pages = {1698--1711},
}

@inproceedings{winstein_stochastic_2013,
	title = {Stochastic {Forecasts} {Achieve} {High} {Throughput} and {Low} {Delay} over {Cellular} {Networks}},
	booktitle = {Proc. of the {USENIX} {Symposium} on {Networked} {Systems} {Design} and {Implementation} ({NSDI})},
	author = {Winstein, Keith and Sivaraman, Anirudh and Balakrishnan, Hari},
	year = {2013},
}

@inproceedings{netravali_mahimahi_2015,
	title = {Mahimahi: {Accurate} {Record}-and-{Replay} for \{{HTTP}\}},
	booktitle = {Proc. of the {USENIX} {Annual} {Technical} {Conference} ({ATC})},
	author = {Netravali, Ravi and Sivaraman, Anirudh and Das, Somak and Goyal, Ameesh and Winstein, Keith and Mickens, James and Balakrishnan, Hari},
	year = {2015},
	pages = {417--429},
}

@misc{gutman_moonlight_2023,
	title = {Moonlight {Mobile} {Cloud} {Gaming} {Client}},
	url = {https://moonlight-stream.org/},
	author = {Gutman, Cameron and Waxemberg, Diego and Neyer, Aaron and Bergeron, Michelle and Hennessy, Andrew and Campbell, Aidan},
	year = {2023},
}

@misc{noauthor_dash_nodate,
	title = {{DASH} {Industry} {Forum}},
	howpublished = {\href{https://reference.dashif.org/dash.js/latest/samples/index.html}{reference.dashif.org}},
    key = {DASH}
}

@inproceedings{yan_pantheon_2018,
	title = {Pantheon: the training ground for {Internet} congestion-control research},
	booktitle = {{USENIX} {ATC}},
	author = {Yan, Francis Y and Ma, Jestin and Hill, Greg D and Raghavan, Deepti and Wahby, Riad S and Levis, Philip and Winstein, Keith},
	year = {2018},
}

@misc{lizardbyte_sunshine_2023,
	title = {Sunshine: a self-hosted game stream host for {Moonlight}.},
	url = {https://github.com/LizardByte/Sunshine/tree/v0.21.0},
	author = {{LizardByte}},
	year = {2023},
}

@misc{system_srsran_2023,
	title = {{srsRAN} {Project}: {Open} {Source} {RAN}},
	url = {https://docs.srsran.com/projects/project/en/latest/tutorials/source/cotsUE/source/index.html},
	author = {System, Software Radio},
	year = {2023},
}

@INPROCEEDINGS{rubner_emd_1998,
  author={Rubner, Y. and Tomasi, C. and Guibas, L.J.},
  booktitle={Sixth International Conference on Computer Vision (IEEE Cat. No.98CH36271)}, 
  title={A metric for distributions with applications to image databases}, 
  year={1998},
  volume={},
  number={},
  pages={59-66},
  doi={10.1109/ICCV.1998.710701}}

@article{karagkioules_l2a_2022,
    author = {Karagkioules, Theodoros and Paschos, Georgios S. and Liakopoulos, Nikolaos and Fiandrotti, Attilio and Tsilimantos, Dimitrios and Cagnazzo, Marco},
    title = {Online Learning for Adaptive Video Streaming in Mobile Networks},
    year = {2022},
    issue_date = {January 2022},
    publisher = {Association for Computing Machinery},
    address = {New York, NY, USA},
    volume = {18},
    number = {1},
    issn = {1551-6857},
    url = {https://doi.org/10.1145/3460819},
    doi = {10.1145/3460819},
    journal = {ACM Trans. Multimedia Comput. Commun. Appl.},
    month = jan,
    articleno = {2},
    numpages = {22}
}

@inproceedings{lolp_lim_2020,
    author = {Lim, May and Akcay, Mehmet N. and Bentaleb, Abdelhak and Begen, Ali C. and Zimmermann, Roger},
    title = {When they go high, we go low: low-latency live streaming in dash.js with LoL},
    year = {2020},
    isbn = {9781450368452},
    publisher = {Association for Computing Machinery},
    address = {New York, NY, USA},
    url = {https://doi.org/10.1145/3339825.3397043},
    doi = {10.1145/3339825.3397043},
    booktitle = {Proceedings of the 11th ACM Multimedia Systems Conference},
    pages = {321–326},
    numpages = {6},
    location = {Istanbul, Turkey},
    series = {MMSys '20}
}

@inproceedings{li2024geo2sigmap,
  title={Geo2SigMap: High-fidelity RF signal mapping using geographic databases},
  author={Li, Yiming and Li, Zeyu and Gao, Zhihui and Chen, Tingjun},
  booktitle={2024 IEEE International Symposium on Dynamic Spectrum Access Networks (DySPAN)},
  pages={277--285},
  year={2024},
  organization={IEEE}
}

@inproceedings{HemmingerNetworkEW,
  title={Network Emulation with NetEm},
  author={Stephen Hemminger},
  url={https://api.semanticscholar.org/CorpusID:17786091}
}

@misc{chrome_devtools_network_throttling,
  author       = {{Chrome Developers}},
  title        = {Network features reference},
  howpublished = {\href{https://developer.chrome.com/docs/devtools/network/reference#throttling-profile}{developer.chrome.com}},
  note         = {Accessed: 2026-04-22}
}

@article{mishra_NemFi_2021,
    author = {Mishra, Abhishek kumar and Ayoubi, Sara and Grassi, Giulio and Teixeira, Renata},
    title = {NemFi: record-and-replay to emulate WiFi},
    year = {2021},
    issue_date = {July 2021},
    publisher = {Association for Computing Machinery},
    address = {New York, NY, USA},
    volume = {51},
    number = {3},
    issn = {0146-4833},
    url = {https://doi.org/10.1145/3477482.3477484},
    doi = {10.1145/3477482.3477484},
    journal = {SIGCOMM Comput. Commun. Rev.},
    month = jul,
    pages = {2–8},
    numpages = {7}
}

@ARTICLE{xiao_channelgan_2022,
  author={Xiao, Han and Tian, Wenqiang and Liu, Wendong and Shen, Jia},
  journal={IEEE Wireless Communications Letters}, 
  title={ChannelGAN: Deep Learning-Based Channel Modeling and Generating}, 
  year={2022},
  volume={11},
  number={3},
  pages={650-654},
  doi={10.1109/LWC.2021.3140102}
}

@article{goodfellow2014generative,
  title={Generative adversarial nets},
  author={Goodfellow, Ian J and Pouget-Abadie, Jean and Mirza, Mehdi and Xu, Bing and Warde-Farley, David and Ozair, Sherjil and Courville, Aaron and Bengio, Yoshua},
  journal={Advances in neural information processing systems},
  volume={27},
  year={2014}
}

@inproceedings{zhenlin_radiotwin_2025,
  author = {Zhenlin An and Longfei Shangguan and John Kaewell and Philip Pietraski and Kyle Jamieson},
  title = {RadioTwin: A Digital Building Material Twin for Wideband, Cross-link, Cross-band Wireless Channel Prediction},
  year = {2025},
  journal = {IEEE International Symposium on Dynamic Spectrum Access (DySPAN)},
  publisher = {IEEE},
  address = {London, UK},
  doi = {10.1109/DySPAN64764.2025.11115919},
}

@INPROCEEDINGS{9024384,
  author={Mota, Mateus P. and Araujo, Daniel C. and Costa Neto, Francisco Hugo and de Almeida, Andre L. F. and Cavalcanti, F. Rodrigo},
  booktitle={2019 IEEE Globecom Workshops (GC Wkshps)}, 
  title={Adaptive Modulation and Coding Based on Reinforcement Learning for 5G Networks}, 
  year={2019},
  volume={},
  number={},
  pages={1-6},
  doi={10.1109/GCWkshps45667.2019.9024384}}

@misc{you2026simulationrealitypracticaldeep,
      title={From Simulation to Reality: Practical Deep Reinforcement Learning-based Link Adaptation for Cellular Networks}, 
      author={Lizhao You and Nanqing Zhou and Guanglong Pang and Jiajie Huang and Yulin Shao and Liqun Fu},
      year={2026},
      eprint={2603.00689},
      archivePrefix={arXiv},
      primaryClass={cs.NI},
      url={https://arxiv.org/abs/2603.00689}, 
}

@misc{sun2020deepreinforcementlearningbasedschedulingcontiguousresource,
      title={Deep-Reinforcement-Learning-Based Scheduling with Contiguous Resource Allocation for Next-Generation Cellular Systems}, 
      author={Shu Sun and Xiaofeng Li},
      year={2020},
      eprint={2010.11269},
      archivePrefix={arXiv},
      primaryClass={cs.NI},
      url={https://arxiv.org/abs/2010.11269}, 
}

@misc{rezazadeh2023multiagentdeepreinforcementlearning,
      title={A Multi-Agent Deep Reinforcement Learning Approach for RAN Resource Allocation in O-RAN}, 
      author={Farhad Rezazadeh and Lanfranco Zanzi and Francesco Devoti and Sergio Barrachina-Munoz and Engin Zeydan and Xavier Costa-Pérez and Josep Mangues-Bafalluy},
      year={2023},
      eprint={2307.02414},
      archivePrefix={arXiv},
      primaryClass={cs.NI},
      url={https://arxiv.org/abs/2307.02414}, 
}

\appendix


\section{Application Categories and Algorithms}
Here we list the application's categories and their corresponding algorithms for data collection. 

\begin{table}[h]
\small
\begin{tabularx}{\linewidth}{lX}
\toprule
\multicolumn{1}{c}{\textbf{Application}}      & \multicolumn{1}{c}{\textbf{Setup}} \\ \midrule
Iperf3    & Random congestion control (CUBIC, BBR, \textit{etc.}, 15 in total) from Linux kernel with random pacing rate [15, 400] Mbps.  \\
Video on demand    & Dash.js\cite{noauthor_dash_nodate} serving Big Buck Bunny with its default ABR algorithm.  \\ 
Video conferencing    & WebRTC\cite{blum_webrtc_2021} peer-to-peer video conferencing with the default Google Congestion Control (GCC).   \\ 
Web-page loading    & Web-page content download (\texttt{wget}) and random wait time in between.\\ 
Cloud gaming    & Sunshine\cite{lizardbyte_sunshine_2023} and Moonlight\cite{gutman_moonlight_2023} with random resolution and FPS combinations.   \\ \bottomrule
\end{tabularx}
\caption{Applications used during data collection.} \label{tab:data_app}
\end{table}

\section{Cell Status}\label{appx:cell_status}
\begin{figure}
    \centering
    \subfigure[User number distribution.]{\includegraphics[width=0.475\linewidth]{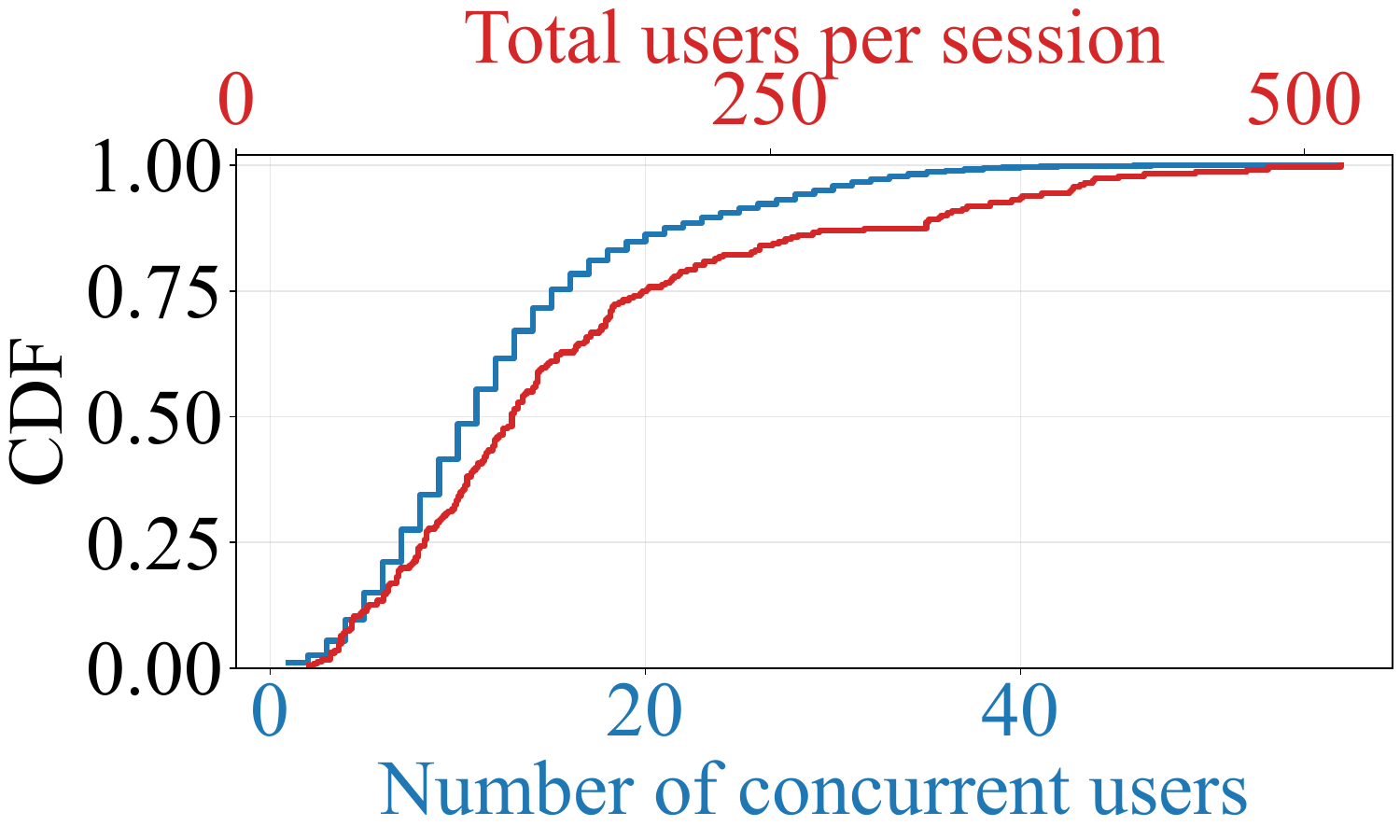}\label{fig:cell-user-count}}
    \hfill
    \subfigure[MCS index distribution.]{\includegraphics[width=0.475\linewidth]{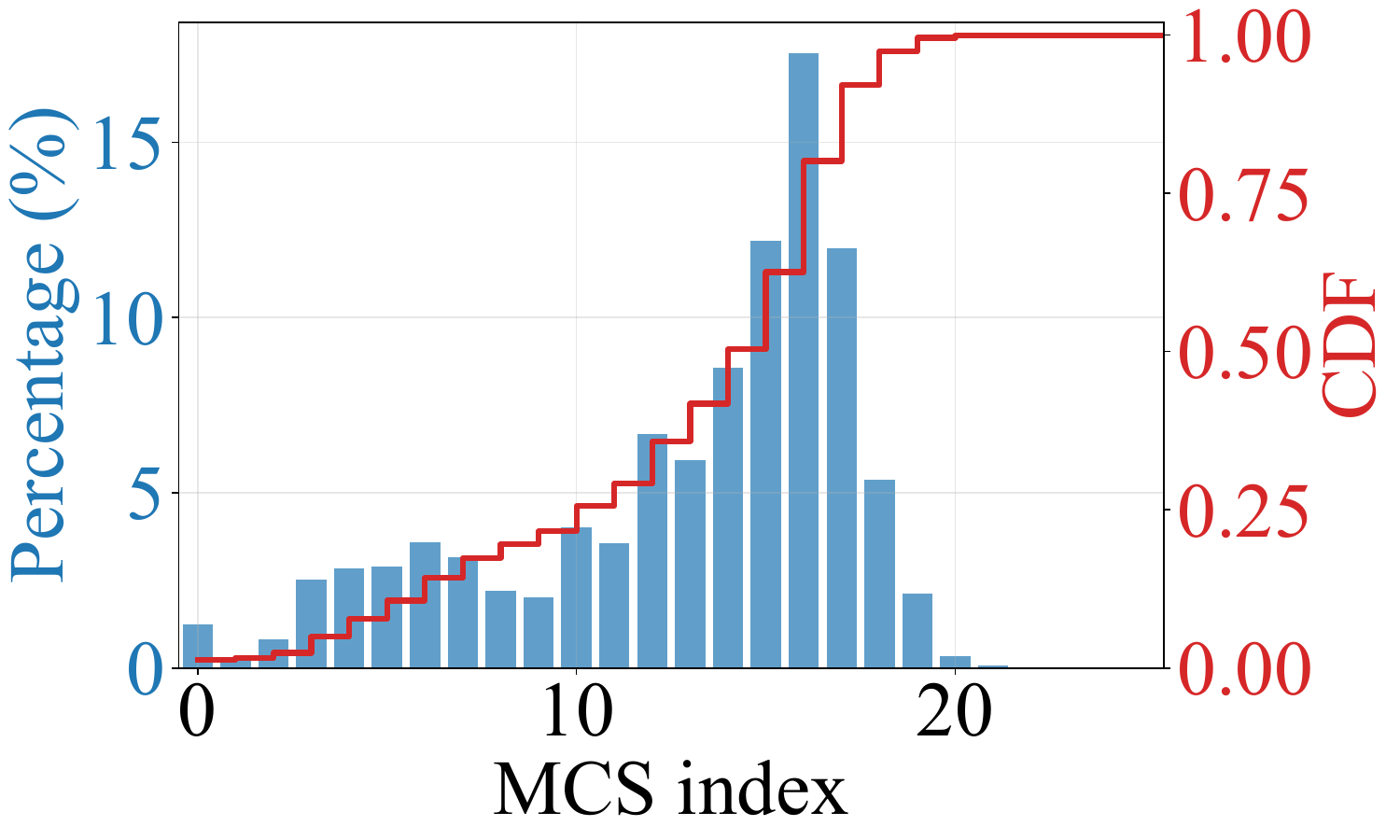}\label{fig:mcs-dist}}
    \caption{User number and MCS distribution in our dataset.}
    \label{fig:cellular-status} 
\end{figure}

Here we summarize the RAN and user status when collecting the data. 
\cref{fig:cell-user-count} shows the concurrent user number distribution over a 1 second window (blue) and the total unique user number throughout the session (red).
The median unique user number across the one second and the whole session are 11 and 129. 
\cref{fig:mcs-dist} shows the distribution of the MCS indexes during our data collection, where the median is 14, and most of the indexes are within 0 to 20.

\section{Copa's Performance}\label{appx:copa}
\begin{figure}
    \centering
    \includegraphics[width=\linewidth]{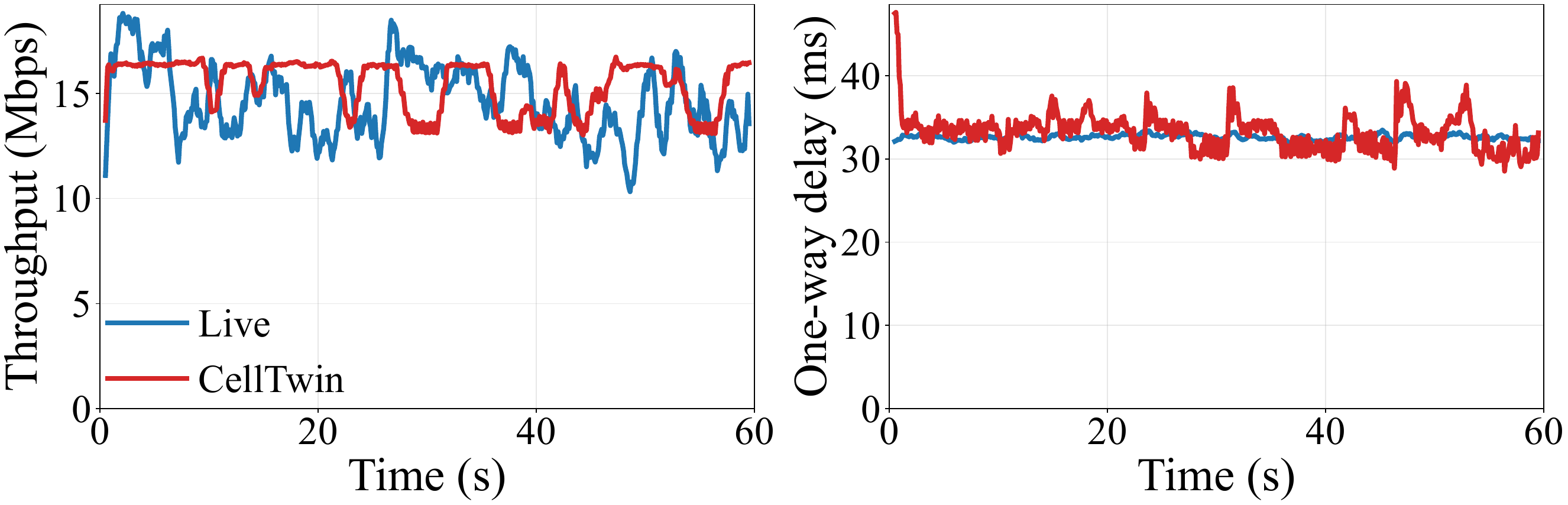}
    \caption{Copa congestion control scheme's performance in live 5G network and \sysname{} emulation.}
    \label{fig:eval_copa}
\end{figure}

Here, we analyze the performance of Copa \cite{arun_copa_2018} as a counterpart case study of the discussion in \S\ref{s:motivation:failure}.
We run the Copa's customized sender and receiver program in live 5G network and \sysname{} emulation. 
\cref{fig:eval_copa} shows the time-series comparison between these two evaluations, where Copa achieves the similar throughput and one-way delay as in the live 5G network, exhibiting significant reduction over the emulation inaccuracy, compared to the time-series result shown in \cref{fig:causality_copa}.

\section{Mobility Evaluation}\label{appx:mobility}

\begin{figure}
    \subfigure[Web page load time.]{\includegraphics[width=0.475\linewidth]{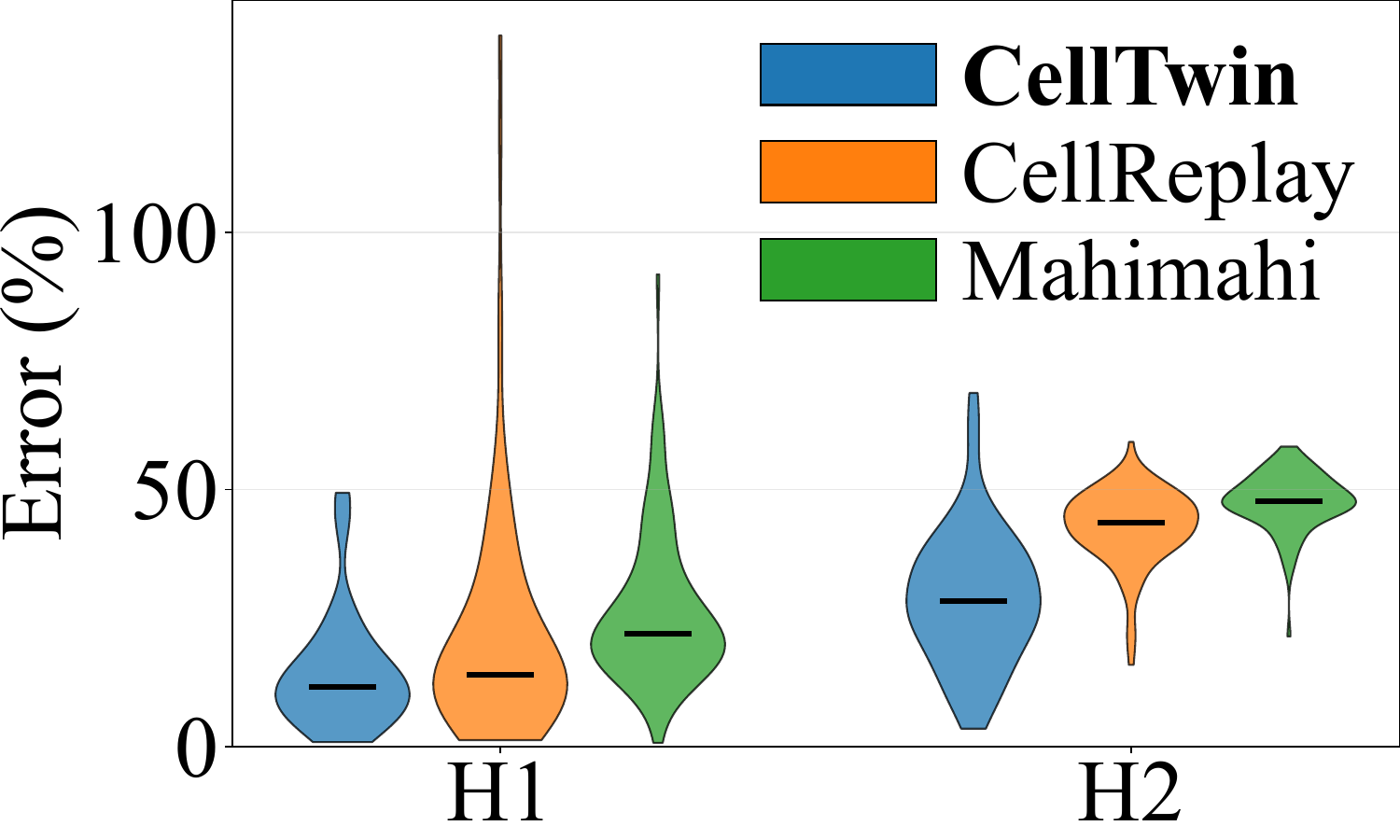}\label{fig:mobile-webpage}}
    \subfigure[VoD bit rate selection.]{\includegraphics[width=0.475\linewidth]{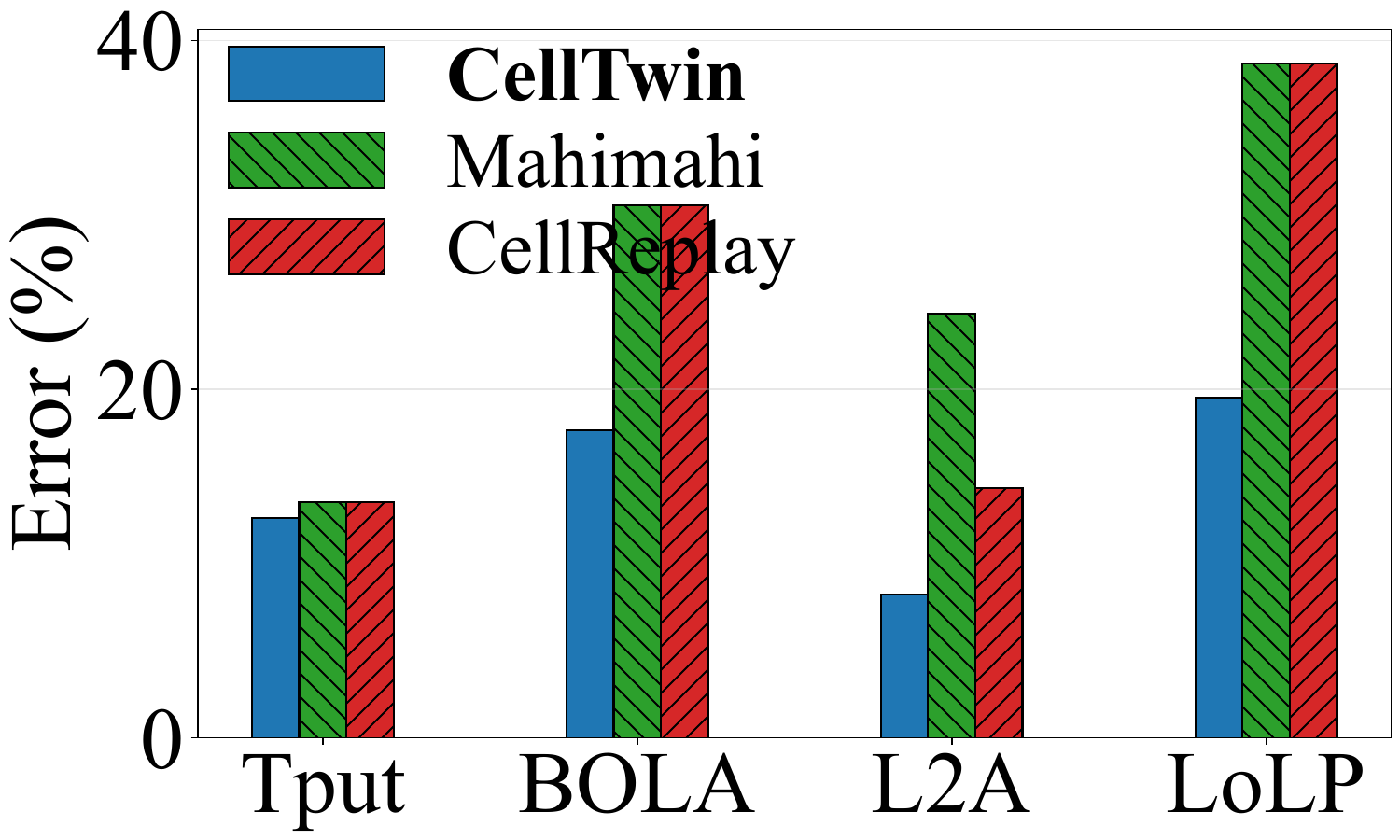}\label{fig:mobile-vod}}
    \subfigure[WebRTC encoder target bit rate.]{\includegraphics[width=0.475\linewidth]{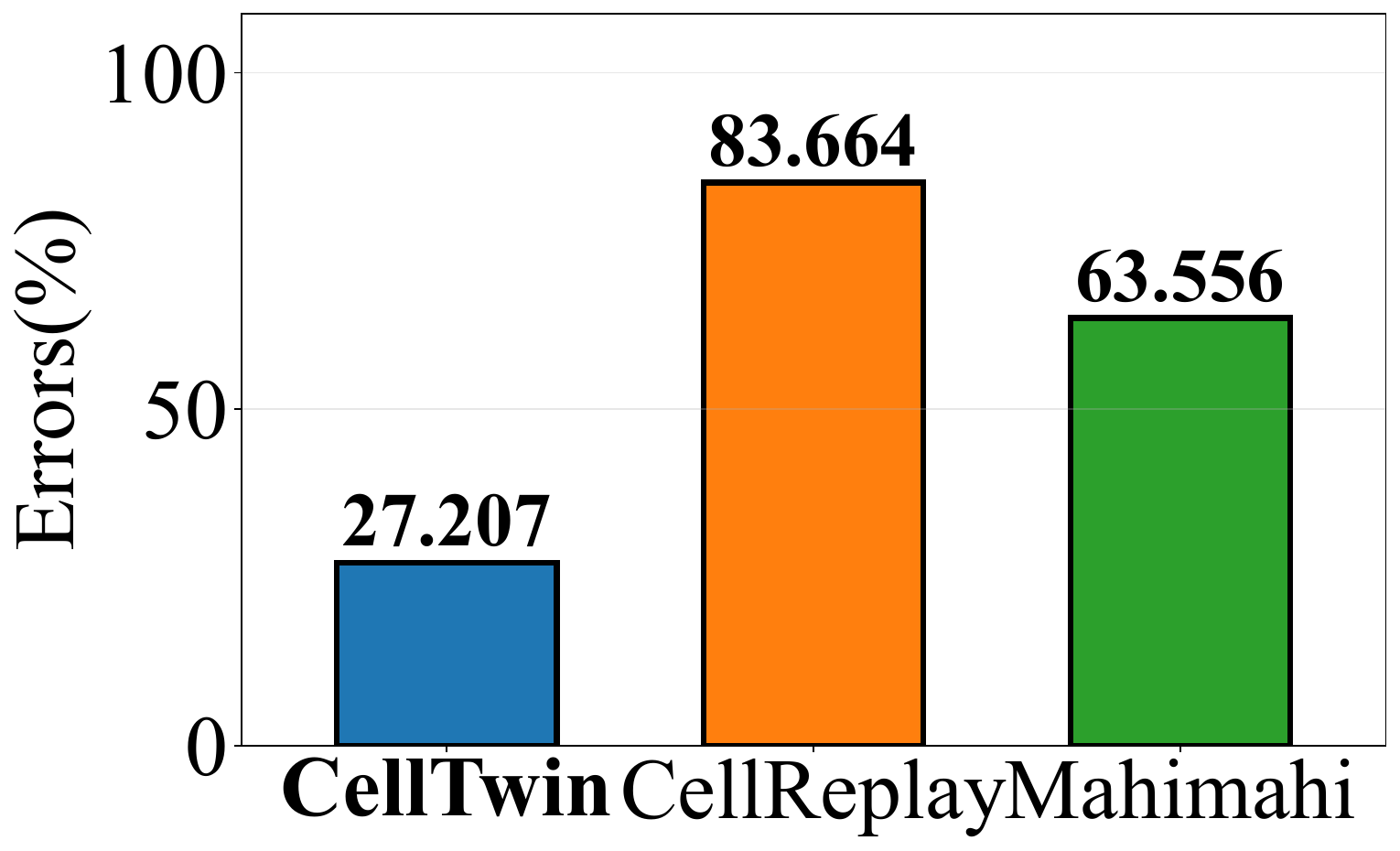}\label{fig:mobile-webrtc}}
    \hfill
    \subfigure[Cloud gaming delay.]{\includegraphics[width=0.475\linewidth]{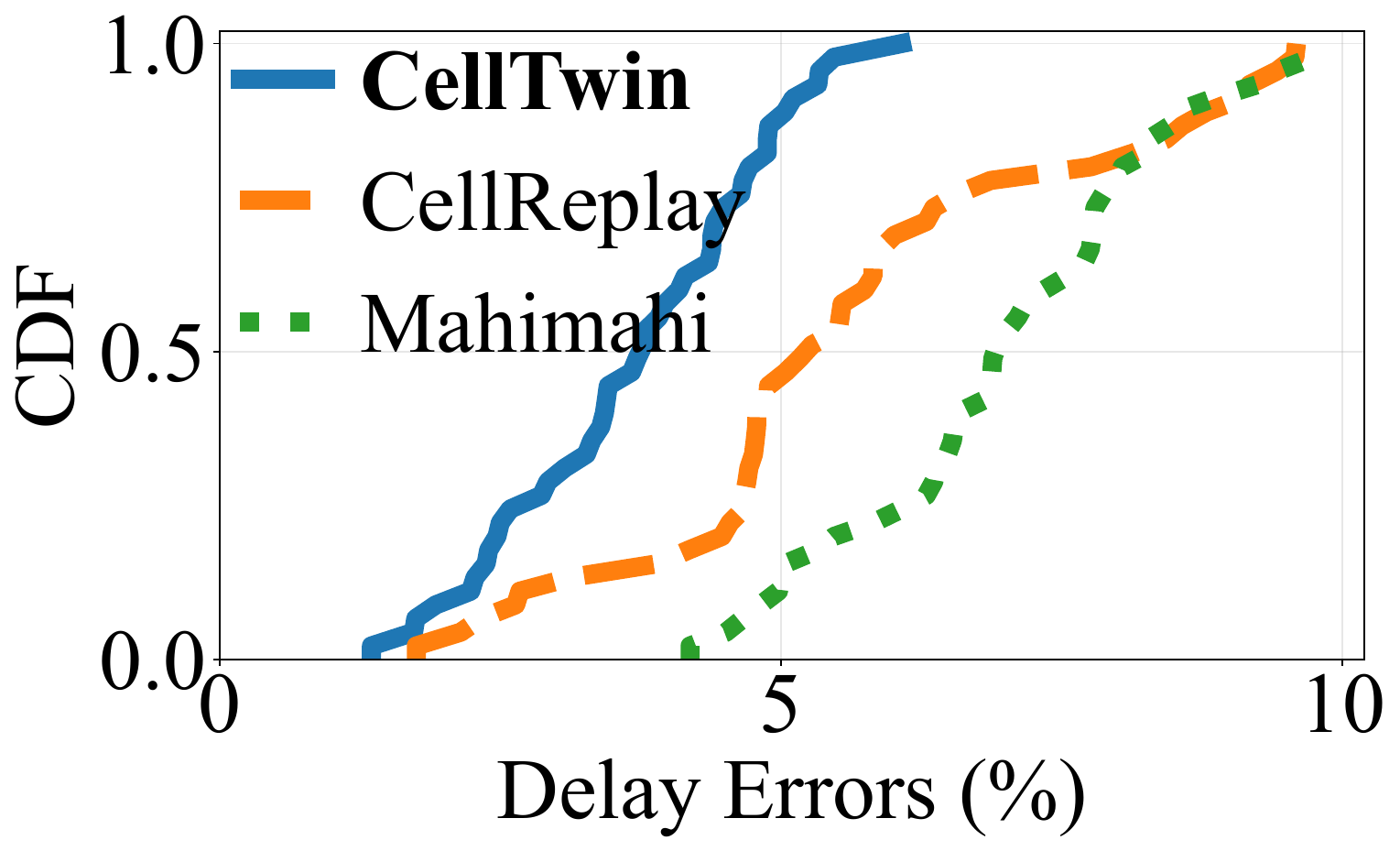}\label{fig:mobile-cg}}
    \caption{Emulation error evaluation results under indoor walking user mobility scenario.}
\end{figure}

We evaluate the emulation accuracy under the indoor walking mobility scenario.
In this evaluation, a user carries the mobile devices and walks following a fixed indoor trajectory. 
During the walk, we run the application first with the live 5G network and collect the CQI and MCS traces for \sysname{} with CellNinja. 
After that we run CellReplay and Mahimahi trace collection program while the user is walking.
With the collected traces, we run all three emulators and compare the results with the live runs through the distribution errors.
We follow the same set of applications as in \S\ref{s:eval:single_user} -- web page loading with HTTP/1.1 and HTTP/2, VoD application, WebRTC peer-to-peer call and cloud gaming.
For each application scenario, we repeat the evaluation for five times and aggregate the results to report.
\cref{fig:mobile-webpage} shows the evaluation results for web page load time for the same 94 web pages, where \sysname{} achieves the lowest emulation errors compared with CellReplay and Mahimahi, where both offer a lower web page load time compared to the live network.
\cref{fig:mobile-vod} shows the bit rate selection in VoD emulation results in the mobility scenario across 4 common ABR algorithms, and \sysname{} maintains an error below 20\%, while CellReplay and Mahimahi yield higher emulation errors.
For WebRTC shown in \cref{fig:mobile-webrtc}, \sysname{} achieves a 27\% bit rate selection emulation error, while CellReplay and Mahimahi's emulation errors exceed 60\%.
As for cloud gaming, all three emulators have low emulation errors, similar as in the static emulation in \S\ref{s:eval:single_user} with the median delay emulation errors being 3.5\%, 5.0\% and 7.2\%, for \sysname{}, CellReplay and Mahimahi, respectively.

\section{Inference Time}
Here we evaluate the inference time of \schedulermodel{}, which generates the real-time results for the emulation system. 
We measure its inference time in the docker container on a desktop machine with 64GB memory, Intel i7-13700K CPU and Nvidia RTX 4060 GPU, deployed with LibTorch-CPU/GPU and TensorRT in the C++ program. 
\cref{tab:inference_time} shows the results, the total TensorRT median inference time is $456.3\ \mu s$ and produces the results for the next 10 slots (5 ms), which fits well into real-time operation.  

\begin{table}[]
\small
\begin{tabular}{cccc}
\hline
                & TensorRT+GPU      & LibTorch+GPU       & LibTorch+CPU       \\ \hline
\textbf{MCS}    & $47.4\ (598.8)$   & $175.5\ (756.3)$   & $199.2\ (377.1)$   \\
\textbf{PRB}    & $408.9\ (987.4)$  & $496.1\ (1134.7)$  & $2690.6\ (3697.7)$ \\ 
\textbf{Total}  & $456.3\ (1586.2)$ & $671.6\ (1891.0)$  & $2889.8\ (4074.8)$ \\ \hline
\end{tabular}
\caption{\sysname{}'s median and (90-th) percentile inference time ($\mu s$) on CPU (i7-13700K) and GPU (RTX 4060).} \label{tab:inference_time}
\end{table}

\end{document}